%% file: recon.tex
\documentclass[letterpaper,twocolumn,10pt]{article}

\usepackage{epsfig,endnotes, usenix, epstopdf}
\usepackage[usenames,dvipsnames]{color}
\usepackage[english,plain]{fancyref}
\usepackage{times}
\usepackage{rotating}
\usepackage[labelformat=simple]{subfig}
\usepackage{multirow}
\usepackage{amsfonts,xspace}
\usepackage{url}
\usepackage{graphicx}
\usepackage[sort]{cite}
\usepackage{verbatim}
\usepackage{balance,capt-of}
\usepackage{flushend}
\usepackage{authblk}
\usepackage{tabularx}

\newcommand{\recon}{{\sl ReCon}\xspace}
\newcommand{\meddle}{{\sl Meddle}\xspace}

\newcommand{\ie}{{\it i.e.,}\xspace}
\newcommand{\eg}{{\it e.g.,}\xspace}

\newcommand{\postfigspace}{0em}

\newcommand{\weka}{{\it weka}\xspace}
\newcommand{\revise}[1]{{#1}}

\newcommand{\new}[1]{{#1}}

\newcommand{\squishenum}{
   \begin{enumerate}{}
    { \setlength{\itemsep}{0pt}      \setlength{\parsep}{0pt}
      \setlength{\topsep}{3pt}       \setlength{\partopsep}{0pt}
      \setlength{\leftmargin}{1.5em} \setlength{\labelwidth}{1em}
      \setlength{\labelsep}{0.5em} } }

\newcommand{\squishlist}{
   \begin{list}{$\bullet$}
    { \setlength{\itemsep}{0pt}      \setlength{\parsep}{3pt}
      \setlength{\topsep}{3pt}       \setlength{\partopsep}{0pt}
      \setlength{\leftmargin}{1.5em} \setlength{\labelwidth}{1em}
      \setlength{\labelsep}{0.5em} } }

\newcommand{\squishlisttwo}{
   \begin{list}{$\bullet$}
    { \setlength{\itemsep}{0pt}    \setlength{\parsep}{0pt}
      \setlength{\topsep}{0pt}     \setlength{\partopsep}{0pt}
      \setlength{\leftmargin}{2em} \setlength{\labelwidth}{1.5em}
      \setlength{\labelsep}{0.5em} } }

\newcommand{\squishend}{
    \end{list}  }

\newcommand{\squishenumend}{
	\end{enumerate}	}

\title{ReCon: Revealing and Controlling PII Leaks 
in Mobile Network Traffic \thanks{This document contains light modifications to our MobiSys 2016 version~\cite{recon:mobisys16}. It includes changes to the layout and additional references.}
} 

\author{Jingjing Ren}
\affil{Northeastern University, Boston, USA}
\author{Ashwin Rao}
\affil{University of Helsinki, Helsinki, Finland}
\author{Martina Lindorfer}
\affil{SBA Research, Vienna, Austria}
\author{Arnaud Legout}
\affil{Inria, Sophia Antipolis, France}
\author[1]{David Choffnes}

\begin{document}

{
\maketitle
}
\input{abstract}
\input{intro}

\input{motivation}

\input{design}

\input{evaluation}

\input{wild}
\input{related}

\input{conclusion}

\bibliographystyle{abbrv}
\begin{small}
\bibliography{mobile-platform-unified}
\end{small}
\vspace{0.001pt} 

\end{document}

%% file: abstract.tex
\section*{Abstract}
It is well known that apps running on mobile devices extensively track and leak users' personally identifiable information (PII); however, these users have little visibility into PII leaked through the network traffic generated by their devices, and have poor control over how, when and where that traffic is sent and handled by third parties. 
In this paper, we present the design, implementation, and evaluation of \recon: a cross-platform system that reveals PII leaks and gives users control over them without requiring any special privileges or custom OSes. \recon leverages machine learning to reveal potential PII leaks by inspecting network traffic, and provides a visualization tool to empower users with the ability to control these leaks via blocking or substitution of PII. 
We evaluate \recon's effectiveness with measurements from controlled experiments using leaks from the 100 most popular iOS, Android, and Windows Phone apps, and via an IRB-approved user study with \new{92} participants. We show that \recon is accurate, efficient, and identifies a wider range of PII than previous approaches. 

%% file: intro.tex
\section{Introduction}
There has been a dramatic shift toward using mobile devices such as smartphones and tablets as the primary 
interface to access Internet services. 
Unlike their fixed-line counterparts, these 
devices also offer ubiquitous mobile connectivity and are equipped with a wide array of 
sensors (\eg GPS, camera, and microphone).

This combination of rich sensors and ubiquitous connectivity makes these devices perfect candidates for privacy invasion. 
Apps extensively track users and leak their personally identifiable information (PII)~\cite{egele:pios,grace:exposure, huber:appinspect,book:collusion,vallina-rod:ads}, and users are generally unaware and unable to stop them~\cite{han2012study,consolvo:ubicomp}. Cases of PII leaks  dramatically increased from 13.45\% of apps in 2010 to 49.78\% of apps in 2014, and the vast majority of these leaks occur over IP networks (less than 1\% of apps leak data over SMS)~\cite{andrubis:badgers}.

Previous attempts to address PII leaks face challenges of a \emph{lack of visibility} into network traffic generated by mobile 
devices and the \emph{inability to control} the traffic.
Passively gathered datasets from large mobile ISPs~\cite{vallina-rod:ads,xia:mosaic} provide visibility but give users no control over network flows. Likewise, custom Android extensions that are often integrated in dynamic analysis tools provide control over network flows but measurement visibility is limited to the devices running these custom OSes or apps~\cite{enck:taintdroid}, often requiring warranty-voiding ``jailbreaking.'' Static analysis tools can identify PII leaks based on the content of the code implementing an app, but suffer from imprecision and cannot defend against dynamic code loading at run time.

We argue that improving mobile privacy requires (1) trusted third-party 
systems that enable auditing and control over PII leaks, and (2) a way for such auditors to identify PII leaks. Our key observation is that a 
PII leak must (by definition) occur over the network, so interposing on network traffic is a naturally platform-independent way to detect and mitigate PII leaks.
Based on this insight, we propose a simpler, more effective strategy than previous approaches: interposing on network traffic 
to improve visibility and control for PII leaks. 

Using this approach, we focus on the problem of identifying and mitigating PII leaks at the network level. We describe the design and implementation of a system to address this problem called \recon, which detects PII leaks from network flows alone, presents this information to users, and allows users fine-grained control over which information is sent to third parties. 
We use machine learning and crowdsourcing-based reinforcement to build classifiers that reliably detect PII in network flows, even when we do not know a priori what information is leaked and in what format. To address flows using SSL or obfuscation, we describe techniques that allow our system to detect PII leaks in encrypted flows with user opt in, and adapt to obfuscation.\footnote{We support SSL decryption for controlled experiments and private \recon instances, but disable them in user studies for privacy reasons.}

By operating on network traffic alone, \recon can be deployed in mobile networks~\cite{awazza}, in home networks, in the cloud, or on mobile devices. \recon is currently deployed using VPN tunnels to software middleboxes running on popular cloud platforms, because this allows us to immediately deploy to arbitrary mobile device OSes and ISPs. 

\smallskip
\noindent
Our key contributions are as follows:

\squishlist
\item A study using controlled experiments to demonstrate how PII leaks from iOS, Android, and Windows Phone devices, motivating the need for (and potential effectiveness of) systems that identify PII leaks from network flows. We find extensive leaks of device identifiers ($>50\%$ of the top 100 apps from all 3 OSes), user identifiers ($>14\%$ of top 100 Android/iOS apps), locations (14-26\% of top 100 Android/iOS apps) and even passwords (3 apps) in \emph{plaintext traffic}.  
\item An approach for the detection and extraction of PII leaks from arbitrary network flows, using machine learning informed by extensive ground truth for more than 72,000 flows generated by mobile apps. 
\item A system that enables users to view PII leaks from network flows, provide feedback about relevant leaks, and optionally modify leaks. 
\item An evaluation of our system, showing it is efficient (classification can be done in less than one ms), and that it accurately identifies leaks (with 98.1\% accuracy for the vast majority of flows in our dataset). 
We show that a simple C4.5 Decision Tree (DT) classifier is able to identify PII leaks with accuracy comparable to several ensemble methods atop DTs (AdaBoost, Bagging, and Blending) that take  significantly more processing time (by a factor of 7.24).
\item A comparison with three alternative techniques for detecting PII leaks using information flow analysis. We show that overall \recon finds more PII leaks than all three approaches. Further, \recon 
can leverage information flow analysis techniques to improve its coverage, as we demonstrate in \S\ref{subsec:compareTD}.

\item A characterization of our approach on traffic generated by user devices as part of an IRB-approved user study. We demonstrate that our approach successfully identifies PII leaks (with users providing \new{5,351} labels for PII leaks) and characterize how these users' PII is leaked ``in the wild.'' For example, we find previously unreported sensitive information such as usernames and passwords (\new{21} apps) being leaked in plaintext flows. 

\squishend

In the next section, we motivate our work using the results of controlled experiments identifying extensive information leakage in popular apps. 
We then describe the design (\S\ref{sec:goaldesign}) and implementation (\S\ref{sec:impl}) of \recon. 
\new{We validate our design choices using controlled experiments in \S\ref{sec:eval}
and in \S\ref{sec:wild} we show their relevance ``in the wild'' with a deployment of ReCon using an 
IRB-approved study with 92 participants.}
We discuss related work in \S\ref{sec:related} and conclude in \S\ref{sec:conc}.
The code and data from our controlled experiments are open-source and publicly 
available at:
\begin{center}
\begin{small}{\url{http://recon.meddle.mobi/codeanddata.html}}\end{small}
\end{center}

%% file: motivation.tex
\section{Motivation and Challenges} 
\label{sec:characterize-app}
In this section, we use controlled experiments to measure PII leakage with ground-truth information. 
We find a surprisingly large volume of PII leaks from popular apps from four app stores, particularly in plaintext (unencrypted) 
flows. Based on these results, we identify several core challenges for detecting PII leaks when 
we do not have ground-truth information, \ie for network traffic generated by arbitrary users' devices. In the next 
section, we describe how to automatically infer PII leaks in network flows when the contents of PII is not known in advance. 

\subsection{Definition of PII}
\label{sec:whatpii}
\emph{Personally identifiable information} (PII) is a generic term referring to ``information which can be used to distinguish or trace an individual's identity''~\cite{omb:pii}. These can include geographic locations, unique identifiers, 
phone numbers and other similar data. 

Central to this work is identifying PII leaked by apps over the network. 
In this paper, we define PII to be either (1) {\bf Device Identifiers} specific to a device or OS installation (ICCID, IMEI, IMSI, MAC address, Android ID, Android Advertiser ID, iOS IFA ID, Windows Phone Device ID), (2) {\bf User Identifiers}, which identify the user (name, gender, date of birth, e-mail address, mailing address, relationship status), (3) {\bf Contact Information} (phone numbers, address book information), (4) {\bf Location} (GPS latitude and longitude, zip code), or (5) {\bf Credentials} (username, password). 
This list of PII is informed by information leaks observed in this study.  While this list is not exhaustive, 
we believe it covers most of the PII that concerns users. We will update the list of 
tracked PII as we learn of additional types of PII leaks.

\subsection{Threat Model}
\label{sec:threatmodel}
\revise{
To improve user privacy, we should inform users of any PII that is exposed to eavesdroppers over insecure connections, and any \emph{unnecessary} PII 
exposed to other parties over secure (\ie encrypted) connections. Determining what information is necessary to share remains an open 
problem that we do not solve in this work, so we consider the upper bound of all PII transmitted to other parties. 

Specifically, we define a ``leak'' as any PII, as described in Section \S\ref{sec:whatpii}, that is sent over the network from a device to a first or third party over both secure (\ie HTTPS) and insecure (\ie HTTP) channels. We further define the following two threat scenarios:

\noindent\textbf{Data-exfiltrating apps.} In this scenario, the app developers either directly, or indirectly via advertising and analytics libraries, collect PII from the users' mobile devices, beyond what would be required for the main functionality of the apps. In this work, we do not establish whether a PII leak is required for app functionality; rather, we make all leaks transparent to users so they can decide whether any individual leak is acceptable.

\begin{sloppypar}
\noindent\textbf{Eavesdropping on network traffic.} Here, the adversary learns PII about a user by listening to network traffic that is exposed in plaintext (\eg at an unencrypted wireless access point, or by tapping on wired network traffic). Sensitive information, such as passwords, are sent over insecure channels, leaving the users vulnerable to eavesdropping by this adversary.
\end{sloppypar}

\recon addresses both scenarios by automatically detecting PII leaks in network flows, presenting the detected leaks to users and allowing them to modify or block  leaks. Clearly, some information should never be sent over insecure channels. Thus, whenever \recon detects a security critical leak, such as a password being sent over HTTP, we follow a responsible disclosure procedure and notify the developer.
}

\subsection{Controlled Experiments for Ground Truth}
\label{sec:dataset-contr-exper}
Our goal with controlled experiments is to obtain ground-truth information about network flows generated by apps and devices.
We use this data to identify PII in network flows and to evaluate \recon (\S\ref{sec:eval}).

\noindent\new{\textbf{Experiment setup.}}
We conduct controlled experiments using Android devices (running Android 5.1.1), an iPhone (running iOS 8.4.1) and a Windows Phone (running Windows 8.10.14226.359). 
We start each set of experiments with a factory reset of the device followed by connecting the device to \meddle~\cite{meddle-tech-report}. 
\meddle provides visibility into network traffic through redirection, \ie sending all device traffic to a proxy server using native support for virtual private network (VPN) tunnels. 
Once traffic arrives at the proxy server, we use software middleboxes to intercept and modify the traffic. 
We additionally use SSLsplit~\cite{sslsplit} to decrypt and inspect SSL flows only during our controlled experiments where \emph{no human subject traffic is intercepted}. \revise{Our dataset and the full details of our experiments are available on our project page at \begin{small}\url{http://recon.meddle.mobi/codeanddata.html}\end{small}.}

\noindent\textbf{Manual tests.} 
We manually test the 100 most popular free apps for Android, iOS, and Windows Phone from the \emph{Google Play} store, the iOS \emph{App Store}, and the \emph{Windows Phone Store} on August 9, 2015 as reported by App Annie~\cite{appannie}. 
For each app, we install it, interact with it for up to 5 minutes, and uninstall it. We give apps permission to access to all requested resources (\eg contacts or location).
This allows us to characterize real user interactions with popular apps in a controlled environment.
We enter unique and distinguishable user credentials when interacting with apps to easily extract the corresponding PII from network flows (if they are not obfuscated). 
\new{Specific inputs, such as valid login credentials, e-mail addresses and names, are hard to generate with automated tools~\cite{choudhary2015automated}. Consequently, our manual tests 
allow us to study app behavior and leaks of PII not covered by our automated tests.}

\noindent\textbf{Automated tests.} 
We include fully-automated tests on the 100 Android apps used in the manual tests and also 850 of the top 1,000 Android apps from the free, third-party Android market \emph{AppsApk.com}~\cite{appsapk} that were successfully downloaded and installed on an Android device.\footnote{14 apps appear both in the AppsApk and Google Play stores, but AppsApk hosts significantly older versions.}
\new{We perform this test 
to understand how third-party apps differ from those in the standard \emph{Google Play} store, 
as they are not subject to \emph{Google Play}'s restrictions and vetting process (but can still be installed by users without rooting their phones).}
We automate experiments using \emph{adb} to install each app, connect the device to the \meddle platform, start the app, perform approximately 10,000 actions using \emph{Monkey}~\cite{adbmonkey}, and finally uninstall the app and reboot the device to end any lingering connections. 
We limit the automated tests to Android devices because iOS and Windows do not provide equivalent scripting functionality.

\noindent\textbf{Analysis.} 
\revise{We use \emph{tcpdump}~\cite{tcpdump} to dump raw IP traffic
and  \emph{bro}~\cite{bro-org} to extract the HTTP flows that we consider in this study, }
then we search  for the conspicuous PII that we loaded onto devices and used as input to text fields. 
We classify some of the destinations of PII leaks as \emph{trackers} using a publicly available database of tracker domains~\cite{yoyoads}, and recent research on mobile ads~\cite{hornyack:appfence, leontiadis:mobileads,crussell2014madfraud}.

\begin{table*}[t]    
    \centering
    \begin{small}
    \begin{tabular}{llll|l|l|l|l|l}
              &             &                  &                  & \multicolumn{5}{c}{\sl \# Apps leaking a given PII } \tabularnewline
             &             & {\bf Testing}    & {\bf \# of }  & {\bf Device}      & {\bf User}        & {\bf Contact}       &                 & \tabularnewline
     {\bf OS}&  {\bf Store}& {\bf Technique}  & {\bf Apps}      & {\bf Identifier}  & {\bf Identifier}  & {\bf Information}   &  {\bf Location} & {\bf Credentials} \tabularnewline       
     \hline
iOS & App Store & Manual & 100 & 47 (47.0\%) & 14 (14.0\%) & 2 (2.0\%) & 26 (26.0\%) & 8
 (8.0\%) 
 \tabularnewline
Android  & Google Play & Manual & 100 & 52 (52.0\%) & 15 (15.0\%) & 1 (1.0\%) & 14 (14.0\%) & 7
 (7.0\%) 
 \tabularnewline
Windows & WP Store & Manual & 100 & 55 (55.0\%) & 3 (3.0\%) & 0 (0.0\%) & 8 (8.0\%) & 1
 (1.0\%) 
 \tabularnewline
Android & AppsApk & Automated & 850 & 155 (18.2\%) & 6 (0.7\%) & 8 (0.9\%) & 40 (4.7\%) & 0
 (0.0\%) 
 \tabularnewline
Android & Google Play & Automated & 100 & 52 (52.0\%) & 0 (0.0\%) & 0 (0.0\%) & 6 (6.0\%) & 0
 (0.0\%) 
 \tabularnewline
    \end{tabular}
    \end{small}
    \caption{\textbf{Summary of PII leaked in plaintext (HTTP) by iOS, Android and Windows Phone apps.} {\new{\sl User identifiers and credentials are leaked across all platforms. Popular iOS apps leak location information more often than the popular Android and Windows apps.}}}
    \vspace{\postfigspace}
    \label{tab:pii}
\end{table*}

\subsection{PII Leaked from Popular Apps}
\label{subsec:exptpii}

We use the traffic traces from our controlled experiments to identify how apps leak PII over HTTP and HTTPS. 
For our analysis we focus on the PII listed in \S\ref{sec:whatpii}.
Some of this information may be required for normal app operation; however, sensitive information such as credentials should never travel across the network in plaintext.  

Table~\ref{tab:pii} presents PII leaked by iOS, Android and Windows apps in plaintext. 
Device identifiers, which can be used to track user's behavior, are the PII leaked most frequently by popular apps.
Table~\ref{tab:pii} shows that other PII---user identifiers, contacts, location, and 
\emph{credentials such as username and password}---are also leaked in plaintext. 
Importantly, our manual tests identify important PII not found by automated tests (\eg, Monkey) 
such as user identifiers and credentials. Thus, previous studies based on automation 
underestimate leakage and are insufficient for good coverage of PII leaks. 

\noindent\textbf{Cross-platform app behavior.} 
We observed that the information leaked by an app varied across OSes.
Of the top 100 apps for Android, 16 apps are available on all the three OSes.
Of these 16 apps, 11 apps leaked PII in plaintext on at least one OS: 2 apps leaked PII on all the three OSes, 5 apps leaked PII in exactly one OS, and the remaining 4 apps leaked PII in 2 of the OSes.
A key take-away is that {\em PII analysis based only on one OS does not generalize 
to all OSes.}

\noindent\textbf{Leaks over SSL.}
During our experiments, we observed that PII is also sent over encrypted channels.  
In many cases, this is normal app behavior (\eg sending credentials when logging in 
to a site, or sending GPS data to a navigation app). However, when such information 
leaks to third parties, there is a potential PII leak. We focus on the PII leaked to tracker 
domains~\cite{yoyoads}, and find that 6 iOS apps, 2 Android apps and 1 Windows app\
send PII to trackers over SSL. The vast majority of 
this information is device identifiers, with three cases of username leaks.
While SSL traffic contains a minority of PII leaks, there is clearly still a need to address 
leaks from encrypted flows.  

Our observations are a conservative estimate of PII leakage because we did not attempt to detect obfuscated PII leaks (\eg via salted hashing), 
and several apps used certificate pinning 
(10 iOS, 15 Android, and 7 Windows apps) or did not work with VPNs enabled (4 iOS apps and 1 Android app).\footnote{Details and the complete dataset can be found on our website.}
Our results in \S\ref{subsec:compareTD} indicate that obfuscation is rare today, and our results above show that significant PII leaks are indeed visible in plaintext.

\subsection{Summary and Challenges}
\label{subsec:sumchall}
While the study above trivially revealed significant PII leaks from popular mobile apps, 
several key challenges remain for detecting PII leaks more broadly. 

\noindent\textbf{Detection without knowing PII.} 
A key challenge is how to detect PII when we do not know the contents of PII in advance. One 
strawman solution is to simply block all advertising and tracking sites. However, this is a blunt and 
indiscriminate approach that can disrupt business models supporting free apps. In fact, the 
developers of the top paid iOS app Peace (which blocks all ads) recently withdrew their app from the App Store 
for this reason~\cite{peace}. 

Another strawman solution is to automatically (and/or symbolically) run every app in every app store to determine when PII is leaked. 
This allows us to formulate a regular expression to identify PII leaks from every app regardless of the 
user: we simply replace the PII with a wildcard.

There are several reasons why this is insufficient to identify PII leaks for arbitrary user flows. 
First, it is impractically expensive to run this automation for all apps in every app store, and there are 
\revise{no publicly available} tools for doing this outside of Android. Second, it is difficult (if not impossible) to 
use automation to explore every possible code path that would result in PII leaks, meaning 
this approach would miss significant PII. Third, this approach is incredibly brittle -- if a tracker 
changes the contents of flows leaking PII at all, the regular expression would fail.  

These issues suggest an alternative approach to identifying PII in network flows: use machine 
learning to build a model of PII leaks that accurately identifies them for arbitrary users. 
This would allow us to use a small set of training flows, combined with user feedback about suspected 
PII leaks, to inform the identification of a PII leaks for a large number of apps. 

\new{\noindent\textbf{Encoding and formatting.}
PII leaked over the network can be encoded using Unicode and other techniques like gzip, JSON, and XML, 
so a technique to identify PII in network flows must support a variety of formats. 
In our experience, it is relatively straightforward to extract the encoding for a flow and search for PII using 
this encoding. 
We support the encodings mentioned above, and will add support for others as we encounter them. }

\noindent\textbf{Encryption.}
Flows in the mobile environment increasingly use encryption (often via SSL). Sandvine reports that in 2014 in North American mobile traffic, approximately 12.5\% of upstream bytes use SSL, up from 9.78\% the previous year~\cite{sandvine:soir}. By comparison, 11.8\% of bytes came from HTTP in 2014, down from 14.66\% the previous year. 
A key challenge is how to detect PII leaks in such encrypted flows. \recon identifies PII leaks 
in plaintext network traffic, so it would require access to the original plaintext content 
to work. While getting such access is a challenge orthogonal to this work, we argue that this is 
feasible for a wide range of traffic if users run an SSL proxy on a trusted computer (\eg the user's \revise{home appliance, such as a computer or home gateway}) 
or use recent techniques for mediated access to encrypted traffic~\cite{blindbox,mctls}.

\noindent\textbf{Obfuscation of PII.} The parties leaking PII may use obfuscation to hide their information leaks. 
In our experiments, we found little evidence of this (\S~\ref{subsec:compareTD}). In the future, we 
anticipate combining our approach with static and dynamic analysis techniques to identify how information is 
being obfuscated, and adjust our system to identify the obfuscated PII. For example, using information 
flow analysis, we can reverse-engineer how obfuscation is done (\eg for salted hashing, learn the salt and hash function), 
then use this information when analyzing network traces to identify leaked PII. In the ensuing cat-and-mouse game, 
we envision automating this process of reverse engineering obfuscation.

%% file: design.tex
\section{ReCon Goals and Design}
\label{sec:goaldesign}

The previous section shows that current OSes do not provide sufficient visibility into PII leaks, provide few options 
to control it, and thus significant amounts of potentially sensitive information is exfiltrated from user devices. 
To address this, we built \recon, a tool that detects PII leaks, visualizes how users' information is shared with various sites, 
and allows users to change the shared information (including modifying PII or even blocking connections entirely). 

The high-level goal of our work is to explore the extent to which we can address privacy issues in mobile systems at the network level. The sub-goals of \recon are as follows:

\squishlist
\item Accurately identify PII in network flows, \emph{without} requiring knowledge of users' PII \emph{a priori}.
\item Improve awareness of PII leaks by presenting this information to users.
\item Automatically improve the classification of sensitive PII based on user feedback.
\item Enable users to change these flows by modifying or removing PII.
\squishend

To achieve the first three goals, we determine what PII is leaked in network flows using network trace analysis, machine learning, and user feedback. 
We achieve the last goal by providing users with an interface to block or modify the PII shared over the network. 
This paper focuses on how to address the research challenges in detecting and revealing PII leaks; as part of ongoing work 
outside the scope of this paper, we are investigating other UIs for modifying PII leaks, how to use 
crowdsourcing to help design PII-modifying rules, and how we can use \recon to provide other types of privacy (\eg 
k-anonymity).

\begin{figure}[t]
\centering
\includegraphics[width=\columnwidth]{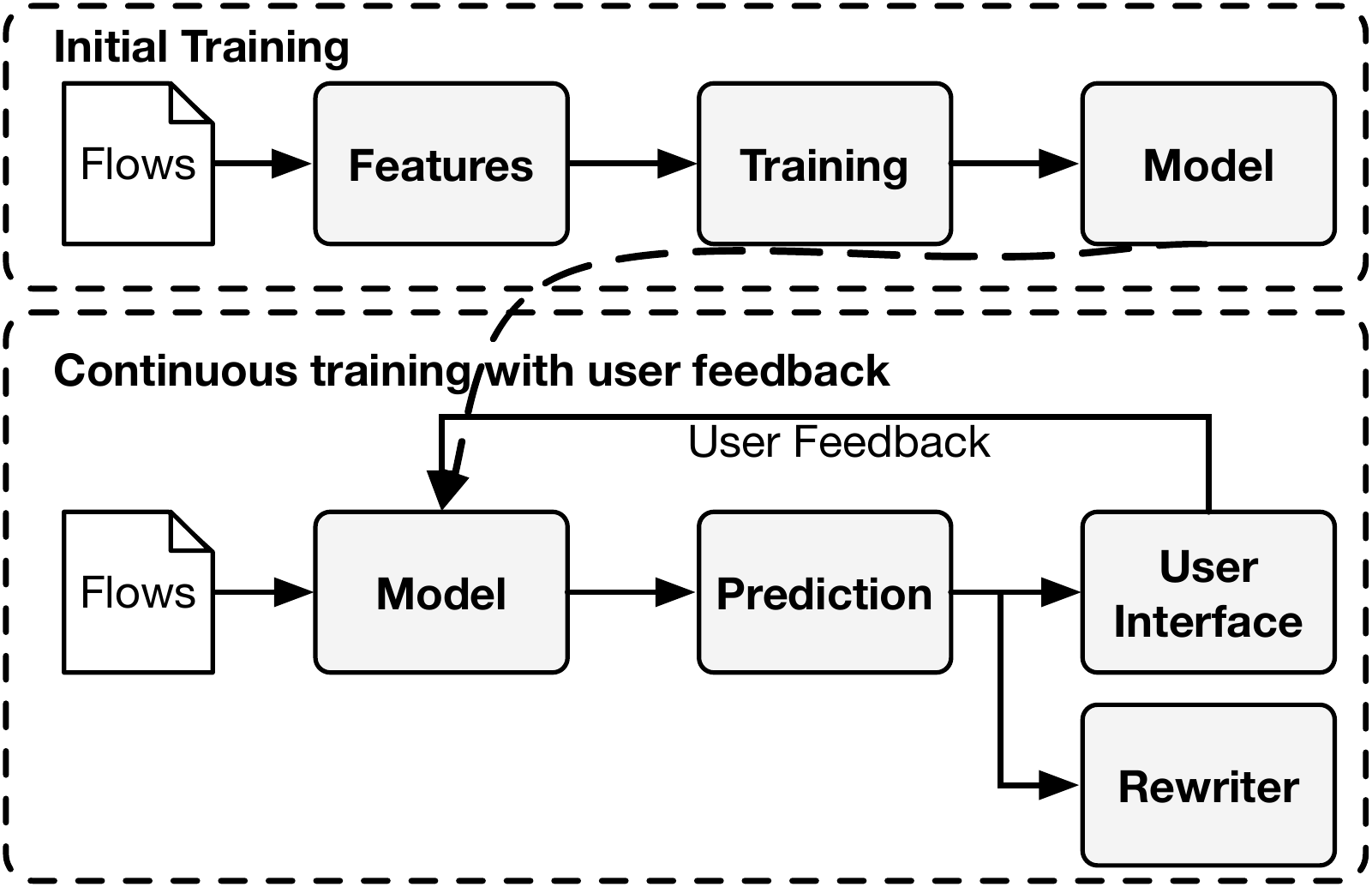}
\caption{\textbf{\recon architecture}. 
{\sl We initially select features and train a model using labeled network flows (top), then use this model to predict whether new network flows are leaking PII. Based on user feedback, we retrain our classifier (bottom). Periodically, we update our classifier with results from new controlled experiments.}}
\label{fig:recon-architecture}
\end{figure}

\begin{figure}[t]
\centering
\vspace{-1em}
\subfloat[PII leaks and actions]{
\includegraphics[width=0.47\columnwidth]{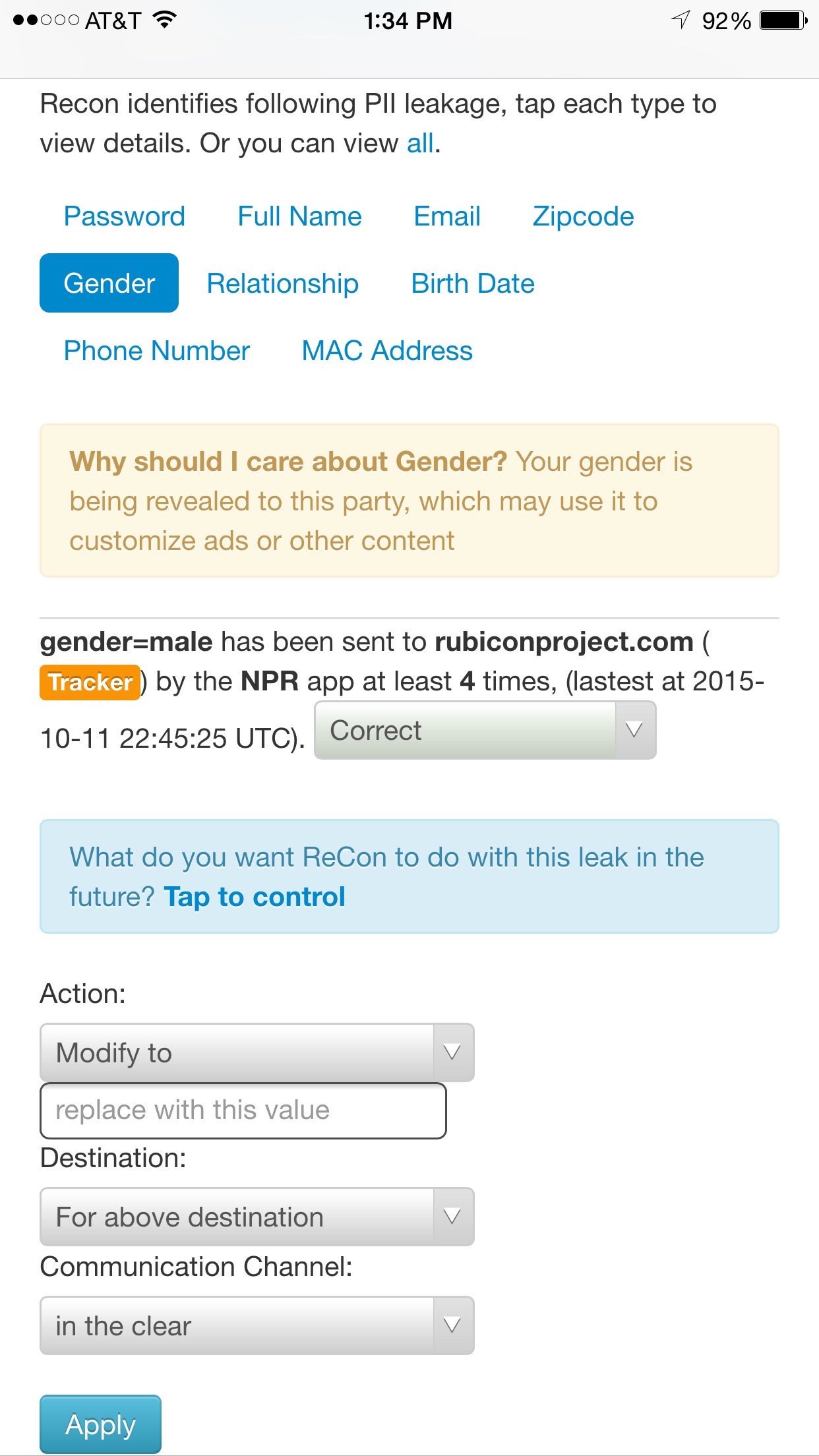}
\label{fig:recon-screenshot-a}
}
\hfill
\subfloat[Map view of location leaks]{
\includegraphics[width=0.47\columnwidth]{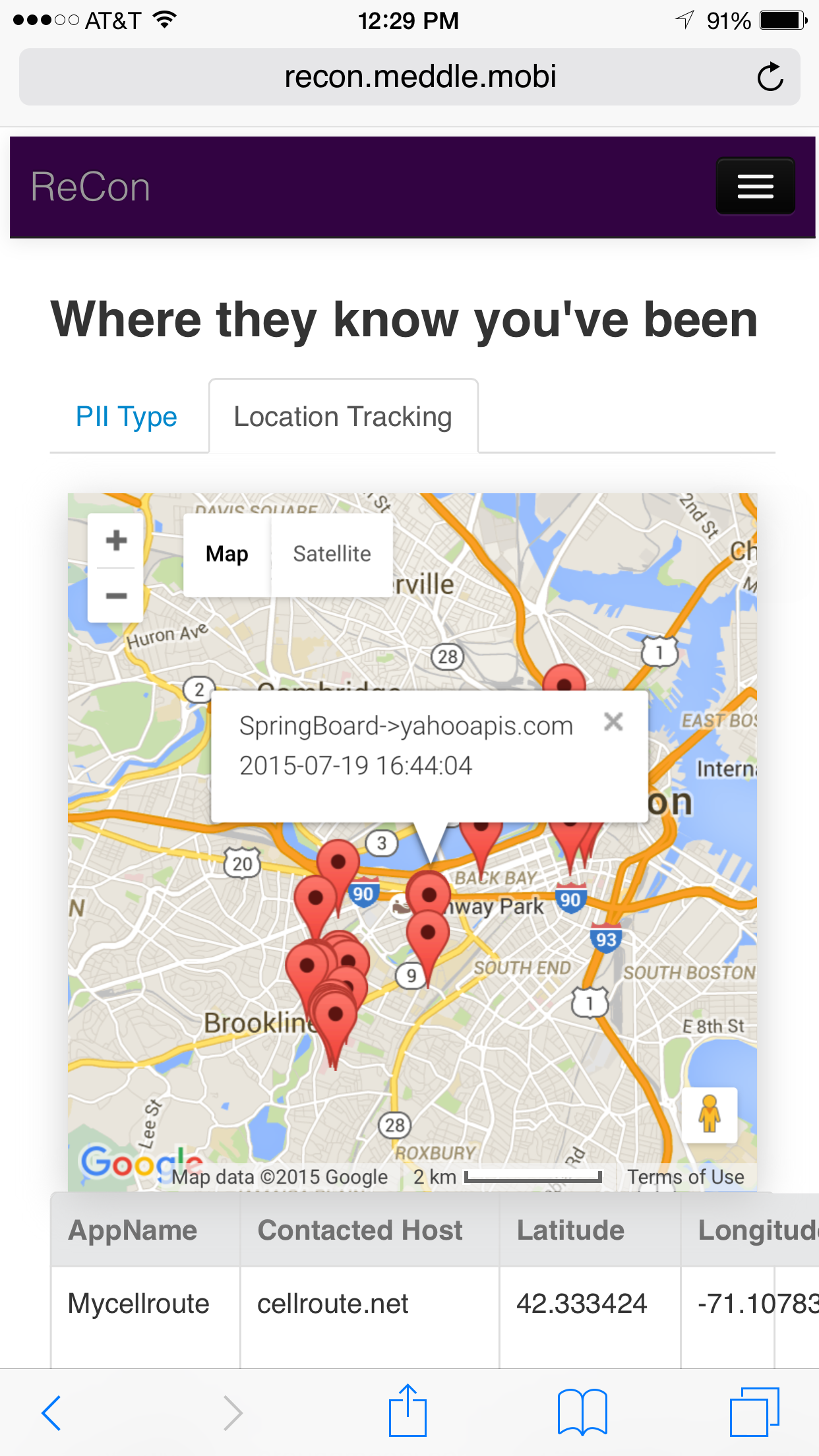}
\label{fig:recon-map}
}
\caption{
{\bf Screen capture of the \recon user interface}.
{\sl Users can view how their PII is leaked, validate the suspected PII leaks, and create custom filters to block or modify leaks.
} }
\vspace{\postfigspace}
\label{fig:recon-screenshot}
\end{figure}

\begin{sloppypar}
Figure~\ref{fig:recon-architecture} presents the architecture of the \recon system. In the ``offline'' phase, we use labeled network traces to determine which features of network flows to use for learning when PII is being leaked, then train a classifier using this data, finally producing a model for predicting whether PII is leaked. When new network flows enter \recon (the ``online'' phase), we use the model to determine whether a flow is leaking PII and present the suspected PII leak to the user via the \recon Web UI (Fig.~\ref{fig:recon-screenshot}). \revise{We currently detect PII as described in the previous section, and will add other PII types as we discover them. Note that our approach can detect any PII that appears in network traffic as long as we obtain labeled examples.}

We collect labels from users (\ie whether our suspected PII is correct) via the UI and integrate the results into our classifier to improve future predictions (left). In addition, \recon supports a map view, where we display the location information that each domain is learning about the user (right). By using a Web interface, \recon users can gain visibility and control into their PII leaks without installing an app.
A demo of \recon is available at \begin{small}\url{http://recon.meddle.mobi/DTL-ReconDemo.mp4}\end{small}. 
\end{sloppypar}

To support control of PII, \recon allows users to tell the system to replace PII with other text (or nothing) for future flows (see the drop-down boxes in Fig.~\ref{fig:recon-screenshot-a}). Users can specify blocking or replacement of PII based on category, domain, or app. This protects users' PII for future network activity, but does not entirely prevent PII from leaking in the first place. To address this, we support \emph{interactive} PII labeling and filtering, using push notifications\footnote{Push notifications require a companion app, and we currently support Android (we plan to release iOS and Windows versions soon).}   or other channels to notify the user of leaks immediately when they are detected (as done in a related study~\cite{balebako:soups13}).

\subsection{Non-Goals} 
\recon is not intended as a blanket replacement for existing approaches to improve privacy in the mobile environment.
For example, information flow analysis~\cite{enck:taintdroid} may identify PII leaks not revealed by \recon. In fact, \recon 
can leverage information flow analysis techniques to improve its coverage, as we demonstrate in \S\ref{subsec:compareTD}.
Importantly, \recon allows us to identify and block unobfuscated PII in network flows from arbitrary devices without requiring OS modifications or taint tracking. 

\revise{The need for access to plaintext traffic is an inherent limitation of our approach. 
We discussed several ways to address encryption and obfuscation of PII in the previous 
section. If these should fail, we can recover plaintext traffic with OS support for access to 
network traffic content as it appears before encryption or obfuscation. Of course, getting such 
support from an OS could be challenging. Alternatively, policymakers such as the FTC could 
intervene by barring developers from using techniques that explicitly eschew auditing tools such 
as \recon, by citing it as a type of ``deceptive business practice'' currently disallowed in the US.  }

\subsection{Deployment Model and User Adoption}
Because \recon needs access only to network traffic to identify and modify PII leaks, it admits a variety of 
deployment models, \eg, in the cloud, in home devices, inside an ISP, or on mobile devices. We are currently 
hosting this service on \meddle in a cloud-based deployment because it provides immediate cross-platform 
support with low overheads~\cite{meddle-tech-report}. 
We are also in discussions with Telefonica to deploy \recon on their Awazza~\cite{awazza} APN proxy, 
which has attracted thousands of users.  

\begin{table*}[ht!]
\centering
\begin{small}
\addtolength{\tabcolsep}{-3pt}
\begin{tabularx}{\linewidth}{c | c | c | X}
\textbf{Section} & \textbf{Topic} & \textbf{Dataset} & \textbf{Key results} \tabularnewline \hline
\ref{subsec:impl-ml} & Implementation & Controlled exp. & Feature extraction and selection, per-domain per-OS classifiers \tabularnewline 
\ref{subsec:pii_extract} & " & Controlled exp. & Automatically identifying PII in flows \tabularnewline 
\ref{subsec:eval-lab} & Evaluation: ML techniques & Controlled exp. & Decision trees provide best trade-off for accuracy/speed, 
per-domain per-OS classifiers outperform general ones, 
feature selection balances accuracy and training time, 
heuristics for PII extraction are accurate \tabularnewline
\ref{subsec:ifa} & Evaluation: IFA comparison & Automated exp. & \recon generally outperforms information flow analysis 
techniques, and can learn new association rules from them to further improve accuracy  \tabularnewline
\ref{sec:wild} & Evaluation: ``in the wild'' & User study & \recon is efficient, users labels confirm accuracy of \recon even for 
apps not previously seen, retraining based on user labels substantially improves accuracy, significant amounts of sensitive 
information is leaked in plaintext from popular apps.  \tabularnewline
\end{tabularx}
\caption{\textbf{Roadmap for key topics covered in \S\ref{sec:impl}, \S\ref{sec:eval} and \S\ref{sec:wild}}. {\new{\sl We train and test our classifier using 10-fold cross-validation, \ie, a random 9/10 samples for training and the remaining 1/10 for testing; we repeat this process 10 times to tune our parameters.}}}
\label{tab:roadmap}
\end{small}
\vspace{-0.5em}
\end{table*}

\subsection{Protecting User Privacy}
An important concern with a \recon user study is privacy. Using an IRB-approved protocol~\cite{meddle-irb-site}, we encrypt and anonymize all captured flows before storing them. We have two deployment models: the first study (approval \#13-08-04) captures all of a subject's Internet traffic and entails in-person, signed informed consent; the second study (approval \#13-11-17) captures only HTTP GET/POST parameters (where most leaks occur) and users consent via an online form. The secret key is stored on a separate secure server and users can delete their data at any time.  
We will make the \recon\ \new{source code} publicly available.
For those who want to run their own \recon instance (\eg if they do not want to participate in our study), 
our system requires only that a user has root on a Linux OS. 
\recon can be deployed 
\recon can be deployed in a single-machine instance on a home computer, as Raspberry Pi plugged into a home router, a dedicated server in an enterprise, \revise{a VM in the cloud, or on the device itself.} One can also selectively route traffic to different \recon instances, \eg, to a cloud instance for HTTP traffic and a trusted home instance \revise{or on-device software such as HayStack~\cite{razaghpanah:2015:haystack}} to decrypt HTTPS connections to identify PII leaked over SSL.

\section{Recon Implementation}
\label{sec:impl}
We now discuss key aspects of our implementation. We then evaluate our design 
decisions in the following section, and finally demonstrate how they hold up ``in the wild'' via a user 
study with 92 participants. Table~\ref{tab:roadmap} presents a roadmap for the remainder of the paper, 
highlighting key design decisions, evaluation criteria, and results. The \recon pipeline begins with parsing network flows, then 
passing each flow to a machine learning classifier for labeling it as containing a PII leak or not.

\subsection{Machine Learning Techniques} 
\label{subsec:impl-ml}
We use the \weka data mining tool~\cite{hall2009weka} to train classifiers that predict PII leaks. We train our classifier by 
extracting relevant features and providing labels for flows that leak PII as described below. \new{Our input dataset is the 
set of labeled flows from our controlled experiments in \S\ref{sec:dataset-contr-exper}. To evaluate our classifiers, we use $k$-fold cross validation, where 
a random $(k-1)/k$ of the flows in our dataset are used to train the classifier, and the remaining 
$1/k$ of the flows are tested for accuracy. This process is repeated $n$ times to understand the stability of our results (see \S\ref{sec:eval}).}

\noindent\textbf{Feature extraction.}
The problem of identifying whether a flow contains PII is similar to the document classification problem,\footnote{Here, network flows are documents and structured data are words.} so we use the ``bag-of-words'' model~\cite{harris54}. We choose certain characters as separators and consider anything between those separators to be words. 
Then for each flow, we produce a vector of binary values where each word that appears in a flow is set to 1, and each word that does not is set to 0. 

A key challenge for feature extraction in network flows is that there is no standard token (\eg whitespace or punctuation) to use for splitting flows into words. For example, a colon (\verb+:+) could be part of a MAC address (\eg \verb+02:00:+\-\verb+00:00:00+), a time-of-day (\eg \verb+11:59+), or JSON data (\eg \verb+username:user007+). Further frustrating attempts to select features, one domain uses ``{\tt=\textgreater}'' as a delimiter (in {\tt username =\textgreater user007}). 
In these cases, there is no single technique that covers all flows. Instead, we use a number of different delimiters ``\verb+,;/(){}[]+" to handle the common case, and treat ambiguous delimiters by inspecting the surrounding content to determine the encoding type based on context (\eg looking at content-en\-coding hints in the HTTP header or whether the content appears in a GET parameter).  

\noindent\textbf{Feature selection.} A simple bag-of-words model produces too many features to be useful for training accurate classifiers that make predictions within milliseconds (to intercept PII leaks in real time). 
To reduce the feature set, we assume that low-frequency words are unlikely to be associated with PII leaks, because when PII does leak, it rarely leaks just once. On the other hand, session keys and other ephemeral identifiers tend to appear in exactly one flow. Based on this intuition, we apply a simple threshold-based filter that removes a feature if its word frequency is too small. We select a reasonable threshold value empirically, by balancing accuracy and classification time for labeled data (discussed in \S\ref{subsec:eval-fs}). \new{To avoid filtering PII leaks that occur rarely in our labeled data, we oversample rarely occurring PII leaks}(so that their number occurrences is greater than the filter threshold). In addition, we randomize PII values (\eg, locations, device IDs) in each flow when training to prevent the classifier from using a PII value as a feature.

While the above filter removes ephemeral identifiers from our feature set, we must also address the problem of words that commonly appear. Several important examples include information typically found in HTTP flows, such as ``content-length:'', ``en-us'', and ``expires''.  We thus add stop-word-based filtering on HTTP flows, where the stop words are determined by term frequency---inverse document frequency (tf-idf). We include only features that have fairly low tf-idf values \new{and that did not appear adjacent to a PII leak in a flow from our controlled experiments}.

\noindent\textbf{Per-domain-and-OS and general classifiers.}
\label{sec:pdao}
We find that PII leaks to the same destination domain use the same (or similar) data encodings to transfer data over the network\new{, but that this encoding 
may differ across different OSes.} 
Based on this 
observation, we build per-domain-and-OS models (one classifier for each \new{[destination domain, OS] pair}) instead of one single general classifier. We identify the domain 
associated with each flow based on the \emph{Host:} parameter in the HTTP header. If this header is not available, we can also identify the domain associated with each IP address by finding the corresponding DNS lookup in packet traces. \new{We identify the OS based on the fact that different OSes use different authentication mechanisms in our VPN, and users tell us in advance which OS they are using.} This improves prediction accuracy because the classifier 
typically needs to learn a small set of association rules. Further, per-domain-and-OS classifiers improve performance in terms of lower-latency predictions \new{(\S\ref{subsec:eval-fs})}, 
important for detecting and intercepting PII leaks in-band.

The above approach works well if there is a sufficiently large sample of labeled data to train to the per-domain per-OS classifier. For domains that do 
not see sufficient traffic, we build a (cross-domain) general classifier. The general classifier tends to have few labeled PII leaks, making it  
susceptible to bias (\eg 5\% of flows in our general classifier are PII leaks). To address this, we use undersampling on negative samples, using 1/10 sampling to randomly choose a subset of available samples. 

\revise{Note that we do not need to train classifiers on every domain in the Internet; rather, we train only on domains contacted by users' traffic. Further, 
we do not need every user to label every PII leak; rather, we need only a small number of labeled instances from a small number of users to identify PII leaks for \emph{all} users whose traffic visits those domains.}

\new{\noindent\textbf{Adapting to PII leaks ``in the wild.''} A key challenge for any ML technique is identifying flows leaking PII that were never seen in controlled experiments. To mitigate this problem, we integrate user feedback from flows that we did identify using one of our classifiers. Specifically, when a user provides feedback that we have correctly identified PII, we can search for that PII in historical flows to identify cases \recon missed due to lack of sufficient training data. Further, we can use these flows to retrain our classifier to successfully catch these instances in future network flows. We discuss the effectiveness of this approach in \S\ref{sec:wild}.}

\revise{Any system that allows user feedback is susceptible to incorrect labels, \eg via user error or Sybil attacks. There are two ways to address this. 
First, we can simply train per-user classifiers, so any erroneous labels only affect the user(s) who provide them. 
Second, we can train system-wide classifiers if we can reliably distinguish good labels from bad ones.
To this end, we envision using existing majority-voting algorithms and/or reputation systems~\cite{jagabathula2014reputation}. }

\subsection{Automatically Extracting PII} 
\label{subsec:pii_extract}
A machine learning classifier indicates whether a flow contains PII, but does not indicate \emph{which content in the flow is a PII leak}. 
The latter information is critical if we want to present users with information about their leaks and allow them to validate the predictions. 

A key challenge for extracting PII is that the key/value pairs used for leaking PII vary across domains and devices; \eg the key ``device\_id'' or ``q'' might each indicate an IMEI value for different domains, but ``q'' \emph{is not always associated with a PII leak}. While we found no solution that perfectly addresses this ambiguity, we developed effective heuristics for identifying ``suspicious'' keys that are likely associated with PII values. 

We use two steps to automatically extract PII leaks from a network flows classified as a leak. 
The first step is based on the relative probability that a suspicious key is associated with a PII leak, calculated as follows:

\vskip5pt
\centerline{$P_{{type,key}}=\frac{K_{{PII}}}{K_{{all}}}$}
\vskip5pt

\noindent where \textit{type} is the PII type (\eg IMEI, e-mail address), \textit{key} is the suspicious key for that \textit{type} of PII, $K_{{PII}}$ is the number of times the key appeared in \revise{flows identified with} PII leaks, and $K_{{all}}$ is the number times the key appeared in all flows. The system looks for suspicious keys that have $P_{{type,key}}$ greater than a threshold. We set this value to an empirically determined value, 0.2, based on finding the best trade-off between false positives (FPs) and true positives (TPs) for our dataset. For users wanting more or less sensitivity, we will make this a configurable threshold in \recon (\eg if a user wants to increase the likelihood of increasing TPs at the potential cost of increased FPs).

In the second step, we use a decision tree classifier, and observe that the root of each tree is likely a key corresponding to a PII value. We thus add these roots to the suspicious key set and assign them a large $P$ value. 

\vspace{1em}
\new{In the next section, we evaluate \recon using controlled experiments on a pre-labeled dataset. This evaluation will only use the initial training phase. Next, we  evaluate \recon in the wild with a user study on our public deployment (\S\ref{sec:wild}). This evaluation will use both the initial training phase and the continuous training phase obtained from real users.}

%% file: evaluation.tex
\section{Evaluation}
\label{sec:eval}

This section evaluates the effectiveness of \recon in terms of accuracy and performance. 
First, we describe our methodology, then we describe the results from controlled experiments 
in terms of classifier accuracy compared to ground truth and to information flow analysis. 
In the next section, we evaluate our system based on results from a user study. 

Our key finding are: 1) we demonstrate that a decision-tree classifier is both 
accurate (99\% overall) and efficient (trains in seconds, predicts in sub-milliseconds); 2) \recon 
identifies more PII than static and dynamic information-flow analysis techniques, and  
can learn from the results of these approaches to improve its coverage of PII leaks. 
Note that this paper focuses on reliably identifying leaks and enabling control, but does not evaluate the control functionality. 

\begin{table*}[ht!]
\centering
\begin{small}
\begin{tabular}{p{4.00cm}|l|l|l|l|l}
                                &         \multicolumn{3}{c|}{\sl Manual tests}        & \multicolumn{2}{c}{\sl Automated tests (Monkey)} \tabularnewline 
                                
{\bf OS (Store)}                        & iOS (App) & Android (Play)	& Windows (WP) &  Android (Play) & Android (AppsApk) \\
 
{\sl Apps tested} & {\sl 100} & {\sl 100} & {\sl 100} & {\sl 100} & {\sl 850}  \\ 
{\sl Apps leaking PII} & {\sl 63} & {\sl 56} & {\sl 61} & {\sl 52} & {\sl 164} \\ \hline
{\bf HTTP flows} & 14683 & 14355 & 12487 & 7186 & 17499  \\
{\bf \ \ \ Leaking PII} & 845 & 1800 & 969 & 1174 & 1776  \\
{\bf Flows to trackers} & 1254 & 1854 & 1253 & 1377 & 5893  \\
{\bf \ \ \ Leaking PII to trackers} & 508 & 567 & 4 & 414 & 649  \\

\end{tabular}
\end{small}
\vspace{0.5em}
\caption{ \textbf{Summary of HTTP flows from controlled experiments.}
{\sl Manual tests generated similar numbers of flows across platforms, but Android leaked proportionately more PII. Collectively, 
our dataset contains more than 6,500 flows with PII leaks.
}}
\label{table-summary}
\end{table*}

\begin{figure*}[ht!]
\centering
\subfloat[Correctly Classified Rate][CCR (x-axis does not start at 0)]{
\includegraphics[width=0.8\columnwidth]{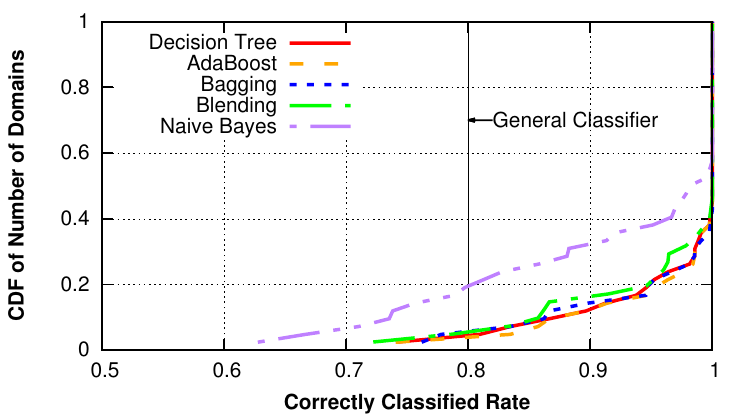}
\label{fig:mlcrr}}
\quad
\subfloat[Area Under Curve ][AUC (x-axis does not start at 0)]{
\includegraphics[width=0.8\columnwidth]{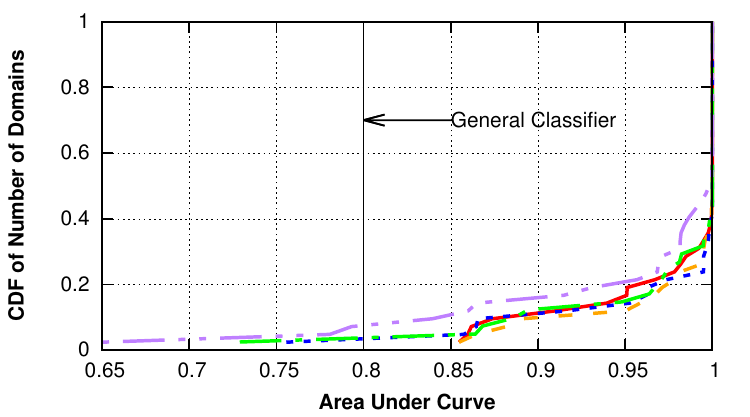}
\label{fig:mlauc}}
\quad
\subfloat[FNR]{
\includegraphics[width=0.8\columnwidth]{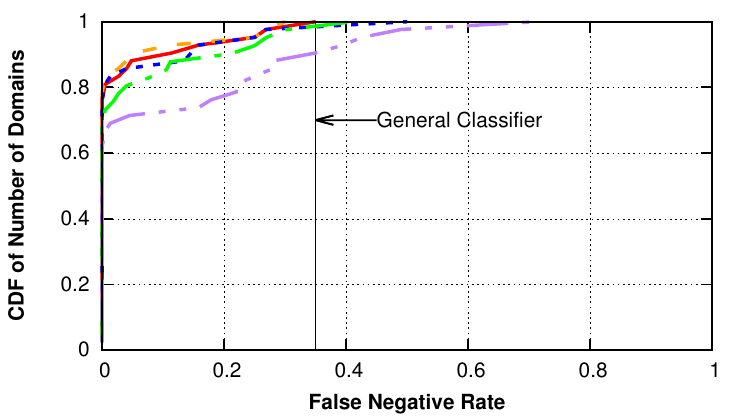}
\label{fig:mlfnr}}
\quad
\subfloat[FPR]{
\includegraphics[width=0.8\columnwidth]{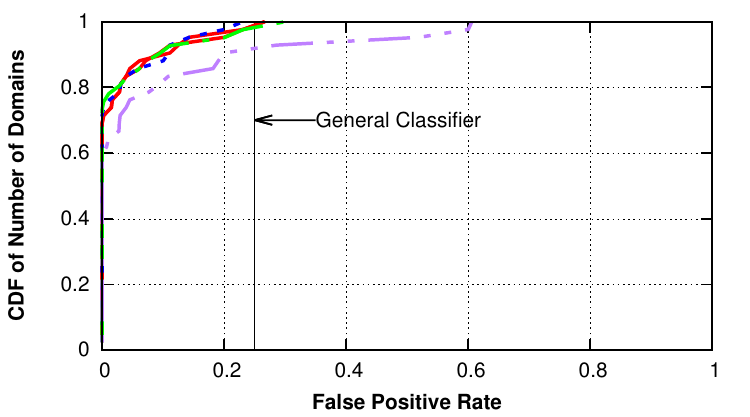}
\label{fig:mlfpr}}
\caption{\textbf{CDF of per-domain-and-OS (PDAO) classifier accuracy, for alternative classification approaches.} 
{\sl For the 42 PDAO
classifiers, DT-based classifiers outperform Naive Bayes, and they exhibit good accuracy (high $CCR$ and $AUC$, low $FPR$ and $FNR$). The vertical line depicts accuracy when using one classifier across all domains, which leads to significantly worse performance.}}
\label{fig:comp-ml}
\end{figure*}

\subsection{Dataset and Methodology}
\label{sec:db}

To evaluate \recon accuracy, we need app-generated traffic and a set of labels indicating which of the corresponding 
flows leak PII. For this analysis, we reuse the data from controlled experiments presented in \S\ref{sec:dataset-contr-exper}; 
Table \ref{table-summary} summarizes this dataset using the number of flows generated by 
the apps, and fraction that leak PII. We identify that more than 6,500 flows leak PII, and a significant 
fraction of those flows leak PII to known trackers. The code and data from our controlled experiments are open-source and publicly 
available at 
 \begin{small}\url{http://recon.meddle.mobi/codeanddata.html}\end{small}.

\new{Recall from \S\ref{subsec:impl-ml} that we use $k$-fold cross-validation to evaluate our 
accuracy by training and testing on different random subsets of our labeled dataset. We 
tried both $k=10$ and $k=5$, and found these values caused only a small difference 
(less than 1\%) in the resulting accuracy.}

We use this labeled dataset to train classifiers and evaluate their effectiveness using the following metrics. We define a positive flow to be one that leaks PII; likewise a negative flow is one that \emph{does not} leak PII. A false positive occurs when a flow does not leak PII but the classifier predicts a PII leak; a false negative occurs when a flow leaks PII but the classifier predicts that it does not. We measure the false positive rate (FPR) and false negative rate (FNR); we also include the following metrics:

\squishlist
\item \textbf{Correctly classified rate (${{CCR}}$)}: the sum of true positive (TP) and true negative (TN) samples divided by the total number of samples. $CCR=(TN+TP)/(TN+TP+FN+FP)$. \\
\emph{A good classifier has a $CCR$ value close to 1.}
\item \textbf{Area under the curve (${{AUC}}$)}: where the curve refers to receiver operating characteristic (ROC). In this approach, the x-axis is the false positive rate and y-axis is the true positive rate (ranging in value from 0 to 1). If the ROC curve is $x=y$ ($AUC=0.5$), then the classification is no better than randomly guessing. \emph{A good classifier has a $AUC$ value near 1.}
\squishend

To evaluate the efficiency of the classifier, we investigate the runtime (in milliseconds) for predicting a PII leak and extracting the suspected PII. We want this value to be significantly lower than typical Internet latencies.

We use the \weka data mining tool to investigate the above metrics for several candidate machine learning approaches to identify a 
technique that is both efficient and accurate. Specifically, we test Naive Bayes, C4.5 Decision Tree (DT) and several ensemble methods atop DTs (AdaBoost, Bagging, and Blending).

\subsection{Lab Experiments}
\label{subsec:eval-lab}
In this section, we evaluate the impact of different implementation decisions and demonstrate the 
overall effectiveness of our adopted approach.

\subsubsection{Machine Learning Approaches}

A key question we must address is which classifier to use. We believe that a DT-based 
classifier is a reasonable choice, because most PII leaks occur in structured data (\ie key/value pairs), 
and a decision tree can naturally represent chained dependencies between these keys and the likelihood 
of leaking PII. 

To evaluate this claim, we tested a variety of classifiers according to the accuracy metrics from the previous section, 
and present the results in Fig.~\ref{fig:comp-ml}. We plot the accuracy using a CDF over the domains that we use 
to build per-domain per-OS classifiers as described in \S\ref{subsec:impl-ml}. The top two graphs (overall accuracy via CCR and AUC), 
show that Naive Bayes has the worst performance, and nearly all the DT-based ensemble methods have 
high CCR and AUC values. (Note that the x-axis does not start at 0.)

Among the ensemble methods, Blending with DTs and k-nearest-neighbor (kNN) yields the highest accuracy; however, 
the resulting accuracy is not significantly better than a simple DT. Importantly, a simple DT takes significantly 
less time to train than ensemble methods. For ensemble methods, the training time largely depends on the number of iterations for training. 
When we set this value to 10, we find that ensemble methods take 7.24 times longer to train than a simple DT on the same dataset. 
Given the significant extra cost with minimal gain in accuracy, we currently use simple DTs.

\begin{figure*}[t]
\centering
\subfloat[a][Simple DT for device identifier]{
\includegraphics[width=0.6\columnwidth]{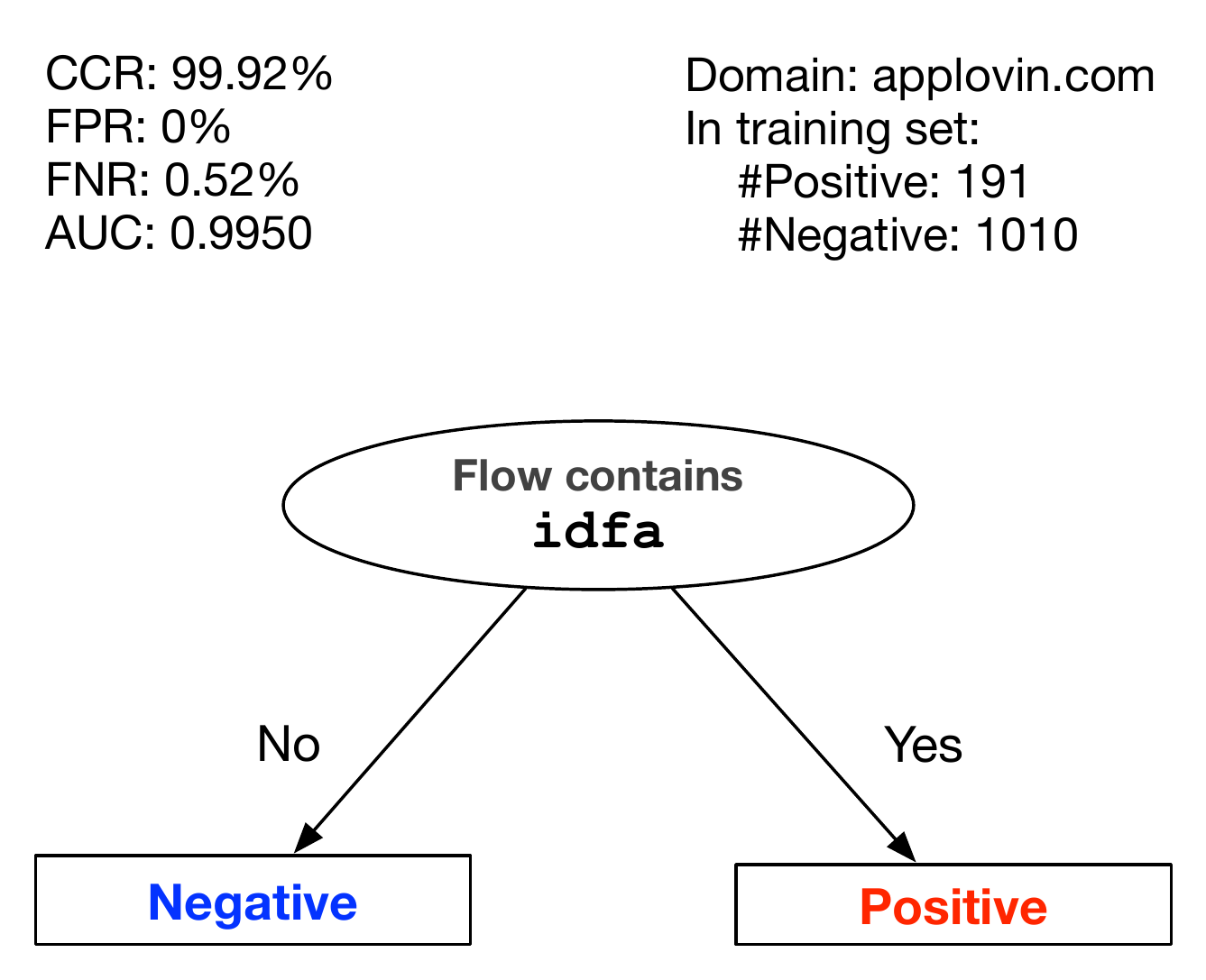}
\label{fig:dt-simple}}
\quad
\subfloat[b][Non-trivial DT for device identifier]{
\includegraphics[width=0.6\columnwidth]{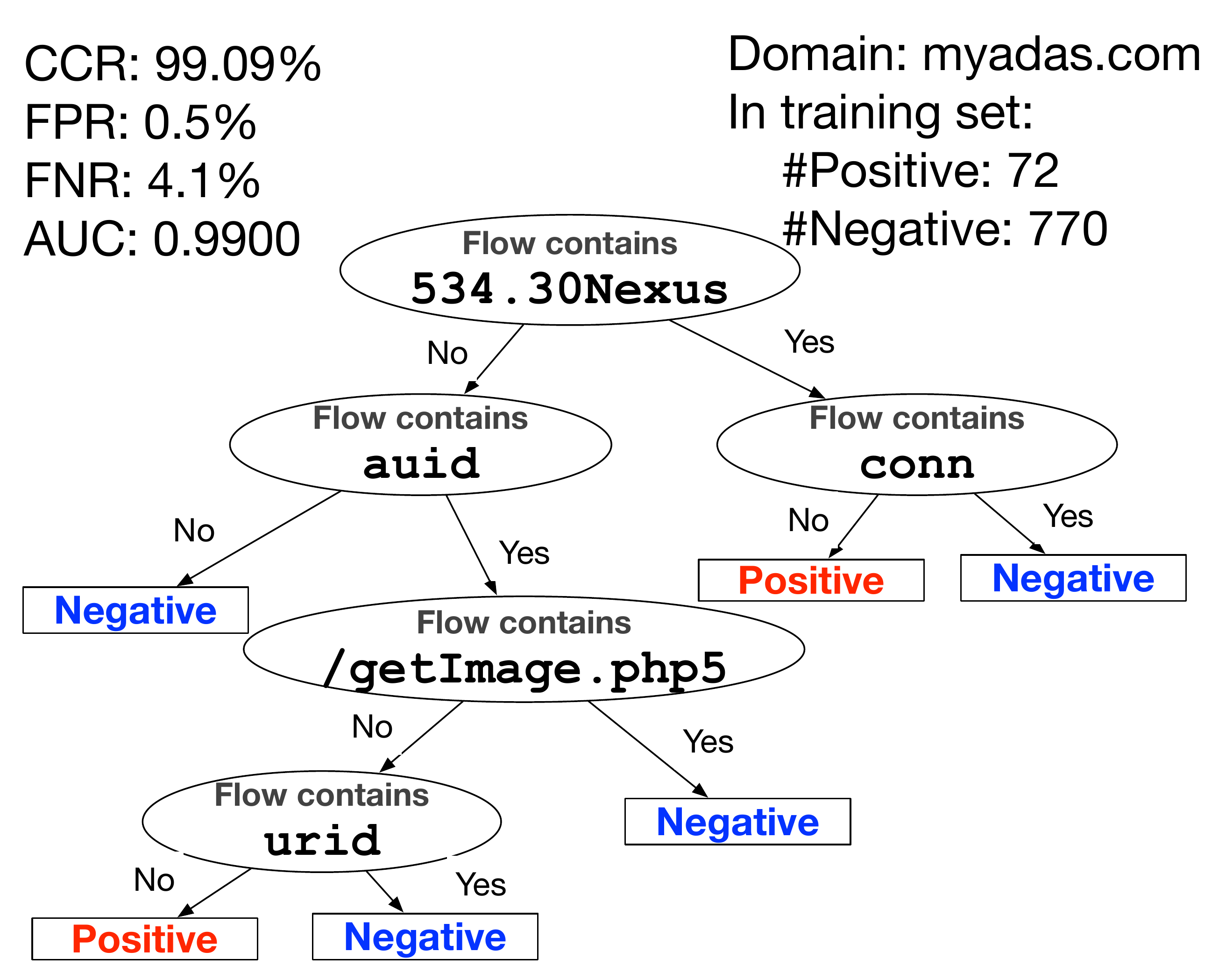}
\label{fig:dt-medium}}
\quad
\subfloat[c][Non-trivial DT for e-mail address]{
\includegraphics[width=0.6\columnwidth]{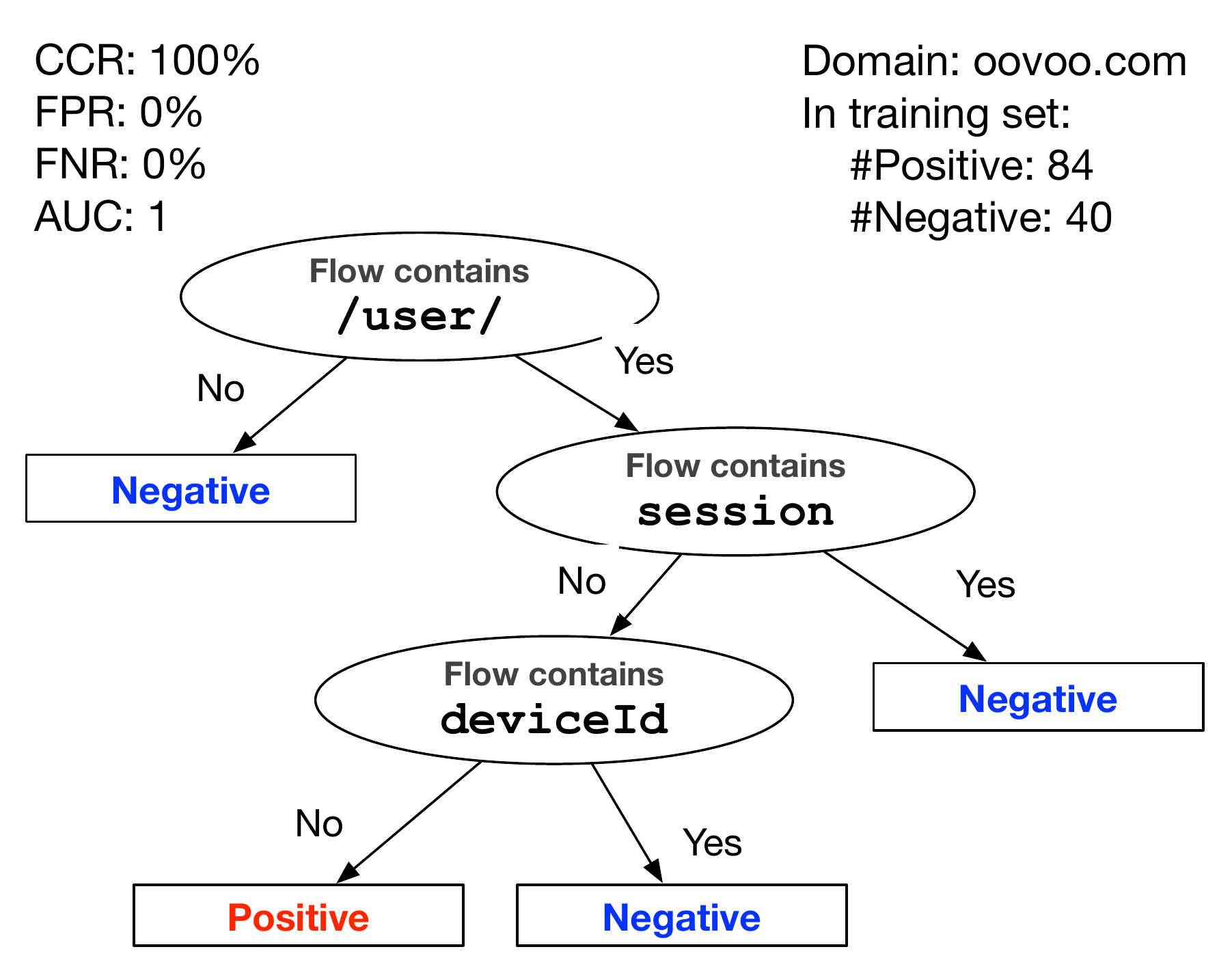}
\label{fig:dt-profile}}
\caption{\textbf{Example decision trees (DTs) for \recon's per-domain per-OS classifiers.} 
{\sl The classifier beings at the root (top) node, and traverses the tree based on whether the term at each node is present. The leaves (boxes) indicate whether there is a PII leak (positive) or not (negative) for each path. The top right of each figure shows the number of positive/negative samples used to train each DT. }}
\label{fig:dt}
\end{figure*}

The bottom figures show that most DT-based classifiers have zero FPs (71.4\%) and FNs (76.2\%)  for the majority of domains. Further, the overall accuracy across all per-domain per-OS classifiers is $>$99\%. The domains with poor accuracy are the trackers {\tt rlcdn.com} and {\tt turn.com}, due to the fact their positive and negative flows are very similar. For example, the key {\tt partner\_uid} is associated both with an Android ID value and another unknown identifier.

To provide intuition as to why DTs work well, and why PII leak detection presents a nontrivial machine-learning problem, we 
include several examples of DTs trained using our data. Some cases of PII leaks are simple: Fig.~\ref{fig:dt-simple} shows that 
Android Advertiser ID is always leaked to the tracker {\tt applovin.com} when the text {\tt idfa} is present in network traffic. Other cases are not trivial, as seen in Fig.~\ref{fig:dt-medium}. Here, we 
find that {\tt auid} is not always associated with an IMEI value, and the DT captures the fact that the IMEI will \emph{not} be present for a {\tt getImage.php5} request if the {\tt urid} is present. 
Finally, Fig.~\ref{fig:dt-profile} gives an example of a non-trivial DT for a different type of PII---e-mail address. Here, the term {\tt email} appears in both positive and negative flows, so this feature cannot be used alone. However, our classifier learns that the leak happens in a {\tt /user/} request when the terms {\tt session} and {\tt deviceId} are not present.\footnote{Note that in this domain {\tt deviceId} is actually used for an app-specific identifier, not a device identifier.} Overall, 62\% of DTs are the simple case (Fig.~\ref{fig:dt-simple}), but more than a third have a depth greater than two, indicating a significant fraction of cases where association rules are nontrivial.

\subsubsection{Per-Domain-and-OS Classifiers}
We now evaluate the impact of using individual per-domain-and-OS (PDAO) classifiers, instead of one general classifier for all flows. We build PDAO classifiers for all domains with greater than 100 samples (\ie labeled flows), at least one of which leaks PII. For the remaining flows, there is insufficient training data to inform a classifier, so we create a general classifier based on the assumption that a significant fraction of the flows use a common structure for leaking PII.\footnote{Note that once \recon acquires sufficient labeled data (\eg from users or controlled experiments) for a destination domain, we create a PDAO classifier.}  

We evaluate the impact of PDAO classifiers on overall accuracy in Figure~\ref{fig:comp-ml}. The vertical lines in the subgraphs represent values for the general classifier, \emph{which is trained using all flows from all domains}. \revise{The figure shows that $>$95\% of the PDAO classifiers have higher accuracy than the general classifier. Further, the high-accuracy PDAO classifiers cover the vast majority of flows in our dataset (91\%). }
Last, training PDAO classifiers is substantially less expensive in terms of runtime: it takes \emph{minutes} to train PDAO classifiers for thousands of flows, but it takes \emph{hours} to train general classifiers for the same flows. 

\subsubsection{Feature Selection}
\label{subsec:eval-fs}

The accuracy of the classifiers described above largely depends on correctly 
identifying the subset of features for training. Further, the training time for classifiers increases 
significantly as the number of features increases, meaning that an efficient classifier requires 
culling of unimportant features. A key challenge in \recon is determining how to 
select such features given the large potential set derived from the bag-of-words approach. 

\begin{figure*}[t]
\centering
\subfloat[\#features changes as threshold changes]{
\includegraphics[width=0.64\columnwidth]{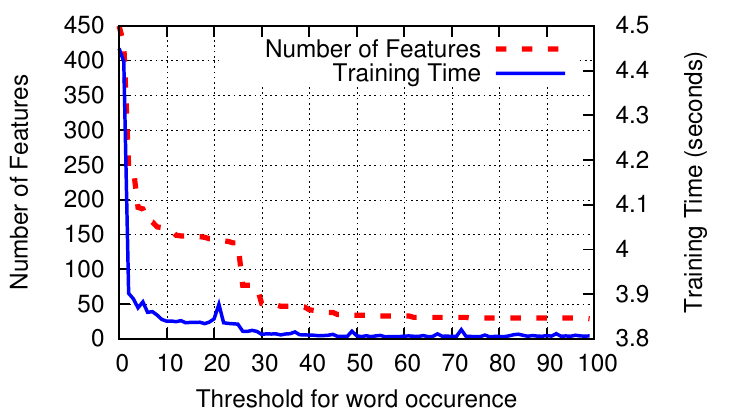}
\label{fig:flurry-threshold}
}
\quad
\subfloat[accuracy and training time over \#features (y-axes do not start at 0)]{
\includegraphics[width=0.64\columnwidth]{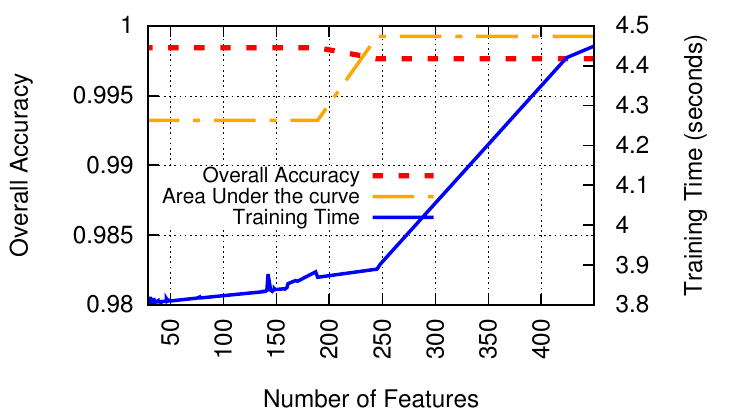}
\label{fig:flurry-feature}
}
\quad
\subfloat[false negative and false positive rate and training time over \#features]{
\includegraphics[width=0.64\columnwidth]{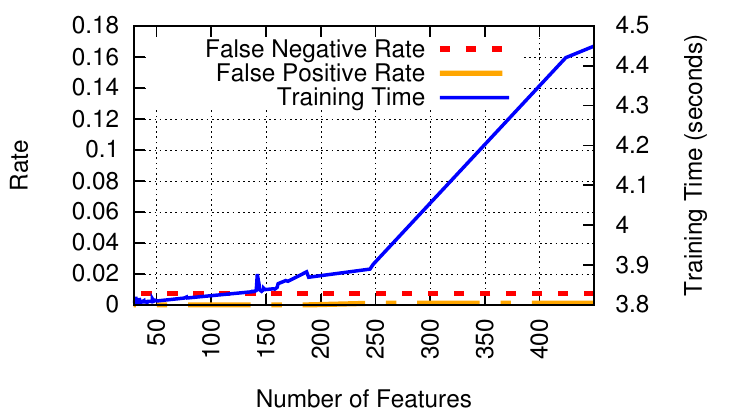}
\label{fig:flurry-feature-fnpr}
}
\caption{\textbf{Feature selection} {\sl for the tracker domain mopub.com. Using $\approx$200 features leads to high accuracy and low training times; however, adding more features increases training time with no benefit to accuracy. }}
\label{fig:selection}
\end{figure*}

We use Figure~\ref{fig:selection} to illustrate this problem and how we address it. Here, we 
focus on statistics for the tracker domain \verb+mopub.com+ (266 flows out of 1,276 leak PII); other domains exhibited similar properties. 

First, we focus on the threshold for including features in our training set. As described in \S~\ref{subsec:impl-ml}, 
we filter out features from words that appear infrequently. Fig.~\ref{fig:flurry-threshold} shows the impact of this 
decision on training time, where the x-axis is the minimum number of appearances for a word to be included 
as a feature, and the y-axis is the time required to train a classifier on the resulting features. The figure 
shows that including all words (threshold = 1) significantly increases training time, but there is a minimal impact on training time 
if the threshold is greater than or equal to 20. The corresponding number of features decreases from 450 to 29 
as the threshold for word occurrence increases from 1 to 99. 

Picking the right number of features is also important for classifier accuracy, as too many features may lead to overfitting and 
too few features may lead to an incomplete model. We evaluate this using Fig.~\ref{fig:flurry-feature}, where the x-axis is 
the number of features, the left y-axis is accuracy (the y-axis does not start at zero), and the right y-axis is training 
time. Even small numbers of features lead to high accuracy for this domain, but increasing the number of 
features beyond 250 does not improve performance (but does increase training time). We see a similar 
effect on the FP rate in Fig.~\ref{fig:flurry-feature-fnpr}.   

While the training time may not seem high in this context, we note that this cost must be incurred 
for each domain and each time we want to update the classifier with user-labeled flows. With potentially thousands of 
flows and labels in a large-scale deployment, such training times can significantly affect the scalability and responsiveness of 
\recon. 

With this in mind, we propose the following strategies for picking threshold values. First, we can use the above 
analysis to find the best threshold, then periodically update this threshold based on new labeled data. Second, we 
can pick a fixed threshold based on the average threshold across all domains (word frequency = 21). We evaluated the impact of these 
two approaches, and found they were nearly identical for our dataset. This suggests that a fixed value is sufficient for 
our dataset, but we propose periodically updating this threshold by performing the above analysis daily or 
weekly as a low-priority background process.

\subsubsection{PII Extraction Strategies}
As discussed in \S~\ref{subsec:pii_extract}, we use two heuristics to identify key/value 
pairs that are likely to leak PII. We use our dataset to evaluate this approach, and find that 
the FP and FN rates are 2.2\% and 3.5\%, respectively. By comparison, a naive approach that 
treats each key/value pair equally yields FP and FN rates of 5.1\% and 18.8\%, respectively. Our approach is thus significantly 
better, and our FP and FN rates are low enough to correctly 
extract PII the vast majority of the time.

\subsection{Comparison with IFA}
\label{subsec:ifa}

\begin{table*}[t]
\centering
\begin{small}
\begin{tabular}{l r | c | c | c | c | c }
\multirow{2}{*}{\bf{Approach}} & \bf{\#apps leaking PII}  & \textbf{Device} & \textbf{User} & \textbf{Contact} & \multirow{2}{*}{\textbf{Location}} & \multirow{2}{*}{\textbf{Credentials}} \tabularnewline 
 & \bf{(\#reports}) & \textbf{Identifier} & \textbf{Identifier} & \textbf{Information} & &  \tabularnewline  \hline
FlowDroid (Static IFA)  & 91 (546)  & 51  (21.52\%)  & 0(-)      & 9 (52.94\%)  & 52 (64.20\%) & $\times$ \tabularnewline 
Andrubis (Dynamic IFA)  & 90 (497)  & 78  (35.46\%)  & $\times$  & 10 (62.50\%) & 3 (3.75\%)   & $\times$ \tabularnewline 
AppAudit (Hybrid IFA)   & 64 (620)  & 57  (24.05\%)  & $\times$  & 3 (17.65\%)  & 4 (4.94\%)   & $\times$ \tabularnewline 
\recon                  & 155 (750) & 145 (61.18\%)  & 6 (100\%) & 4 (23.53\%)  & 29 (35.80\%) & 0 (-)\tabularnewline \hline
Union of all approaches & 278 (750) & 237            & 6         & 17           & 81           & 0
\end{tabular}
\caption{\textbf{Comparison of \recon with information flow analysis (IFA) tools.}  {\sl This comparison is based on automated tests for 750 Android apps (apps from the Google Play and AppsApk dataset for which we observed network flows). We present the number of Android apps detected as leaking PII \revise{(or in the case of FlowDroid, flagged as potentially leaking PII)}, as well as the percentage of leaking apps detected by each tool out of all leaking apps detected by any of the tested tools in each category ($\times$ means the tool does not track that type of information). User credentials were not leaked because our automation tools cannot input them. } 
}
\label{table-infoflow}
\end{small}
\end{table*}

Our labeled dataset in the above analysis may miss PII leaks that are obfuscated 
or otherwise hidden from our analysis. We now evaluate our approach by comparing with one 
that is resilient to such issues: information flow analysis (IFA). We experiment with three IFA 
techniques: (1) static IFA with FlowDroid~\cite{arzt:flowdroid}, 
(2) dynamic IFA with TaintDroid~\cite{enck:taintdroid} (via Andrubis~\cite{andrubis:badgers}), 
and (3) AppAudit~\cite{appaudit}, which uses a combination of both static and approximated dynamic analysis. 
Each of these tools has limitations: some are very resource intensive and some pose 
restrictions on the type of apps they can successfully analyze.

\noindent\textbf{Static IFA.}
FlowDroid detects PII leaks as data flowing between sensitive sources and sinks, which are configured via a list of 
Android API calls. However, the analysis is quite resource intensive: for 
\revise{4.99\%} of apps, our available memory of 8GB was insufficient
for analysis; for 
\revise{17.24\%} of apps the analysis exceeded our analysis timeout of 30 minutes. 
The detected leaks are reported as paths between the API calls. 
\revise{Note that this approach can lead to false positives, since a detected 
leak may never be triggered during app execution.}

\noindent\textbf{Dynamic IFA.}
Andrubis is an app analysis sandbox that uses TaintDroid to identify PII leaks from Android apps during dynamic analysis.
Andrubis installs each app in an emulated Android environment and monitors its behavior for 240 seconds. 
Besides calling all of the app's registered components and simulating common events, 
such as incoming SMS and location changes, it uses Monkey~\cite{adbmonkey} to generate approximately 
8,000 pseudo-random user events. In addition to detailed analysis report 
including all detected data leaks, it also provides the recorded network packet traces.
However, this analysis fails for 
\revise{33.73\%} of apps because they exceed the file size and/or API level limit of Andrubis.

\noindent\textbf{Hybrid IFA.}
AppAudit flags functions that potentially leak PII through static analysis, then performs 
simulated dynamic analysis to filter out candidate functions to confirm PII leaks. It reports leaks 
to the network, file system and through SMS from sources such as the location, contacts and device  
identifiers. The analysis failed for 
\revise{17.33\%} of apps. Note that AppAudit only approximates the 
execution of suspicious functions, and thus does not record any network packet traces.

\noindent\textbf{Methodology and results.} We use the 850 apps 
from \emph{AppsApk.com} and the top 100 apps from \emph{Google Play} from \S\ref{sec:dataset-contr-exper}, and focus on the \revise{750} 
apps that produced network traffic in our experiments. Since static and hybrid IFA approaches do not provide network flows, 
\revise{they only indicate whether an app will potentially leak a certain type of PII. To compare these techniques 
with dynamic analysis,} we base our comparison on the \emph{number of apps} that potentially leak a certain type of PII.
Specifically, we flag an app as leaking a certain type of PII  
if any tool detected \revise{an actual or potential} PII leak in that category (this occurs for \revise{278} apps). 
\revise{We further filtered out cases where dynamic analysis incorrectly flagged a PII leak.}

Table~\ref{table-infoflow} shows the number and percentage of apps that were flagged by FlowDroid, Andrubis, 
AppAudit and \recon. FlowDroid mainly identified \revise{potential} location and phone number leaks, while AppAudit 
mainly identified IMEI leaks. Andrubis performed well in detecting device identifiers (ICCID, IMEI, IMSI) 
and the phone number. \emph{Importantly, \recon identifies more PII leaks overall, and in more categories than 
IFA.}

The above results are encouraging for \recon, and we further investigated mismatches 
between \recon and TaintDroid results, since the latter provides network traces that we 
can process via \recon.
Note, as the authors of TaintDroid themselves acknowledge~\cite{enck:taintdroid}, it may generate false 
positives (particularly for arrays and IMSI values), due to 
propagating taint labels per variable and IPC message. We thus manually inspected flows 
flagged as leaking PII, and discarded cases where the identified PII did not appear in plaintext network flows (\ie false positives). 
Table~\ref{table-taintdroid} shows the results of our analysis, grouped by PII type.

\label{subsec:compareTD}
\begin{table}[t]
\centering
\begin{small}
\addtolength{\tabcolsep}{-1pt}
\begin{tabular}{ll|c|c|c|c|c}
             &                  & \multicolumn{5}{c}{\sl Type of PII being leaked} \tabularnewline
             & {\bf\# leaks}      & {\bf Device}      & {\bf User}        & {\bf Con-} & {\bf Loca-} & {\bf Cred-} \tabularnewline
   & {\bf detected}   & {\bf Id.}         & {\bf Id.}         & {\bf tacts}   & {\bf tion}  & {\bf entials}\tabularnewline       
\hline
\multirow{4}{*}{{\bf \rotatebox{90}{Andrubis}}}
              & plaintext        & 173 & N/A & 10   &  8 & N/A \tabularnewline
              & obfuscated       & 124  & N/A & 16  &  0 & N/A\tabularnewline
              & incorrect        & 140  & N/A & 24  &  6 & N/A \tabularnewline
              & Total      & 457 & N/A & 50   &  14 & N/A\tabularnewline
& & & & & & \\
\multirow{2}{*}{{\bf \rotatebox{90}{ReCon}}}
              & TP    & 146 & 17  & 7   & 35 & 0\tabularnewline
              & FN  & 27  & 0   & 0    & 0  & 0\tabularnewline
\end{tabular}
\end{small}
\caption{\textbf{Comparison with Andrubis (which internally uses TaintDroid), for Android apps only.} 
{\sl Note that this table counts the number of flows leaking PII, not the number of apps. 
TaintDroid has a higher false positive rate than \recon, but catches more device 
identifiers. After retraining \recon with these results, \recon correctly identifies all 
PII leaks. Further, \recon identifies PII leaks that TaintDroid does not. }
}

\label{table-taintdroid}
\vspace{-0.5em}
\end{table}

We use the plaintext leaks identified by Andrubis as ground truth, and evaluate our system by sending the Andrubis network traffic through  \revise{ \recon trained with the pre-labeled dataset described in Section~\S\ref{sec:db}}. The \recon false positive rate was quite low (0.11\%), but the false negative rate was relatively high (15.6\%). The vast majority of false negative flows were Device ID leaks (124/457 are obfuscated and 140/457 are false positive reports from Andrubis). \emph{Importantly, when we retrain \recon's classifier with the Andrubis data, we find that all of the false negatives disappear.} Thus, \recon is \emph{adaptive} in that its accuracy should only improve as we provide it more and diverse sets of labeled data. In the next section we describe results suggesting that we can also use crowdsourcing to provide labeled data. 

\revise{In addition, we can use network traces labeled by IFA to train \recon even in the presence of PII obfuscation. This works because \recon does not search for PII itself, but rather the contextual clues in network traffic that reliably indicate that PII is leaking.}
 
Finally, \recon identified several instances of PII leaks that are not tracked by IFA. These include 
the Android ID, MAC address, user credentials, gender, birthdays, ZIP codes, and e-mail addresses.

%% file: wild.tex
\section{ReCon in the Wild}
\label{sec:wild}

\begin{sloppypar}
We now describe the results of our IRB-approved user study, where participants used \recon 
for at least one week and up to over 200 days, interacted with our system via the UI, and completed a follow-up 
survey. \new{Our study is biased toward flows from the US due to initial recruitment in the Boston area, but includes connections from users in 21 countries in four continents.} While we cannot claim representativeness, 
we can use the user feedback quantitatively, to understand the impact of labeling on 
our classifiers. We also use the study qualitatively, to understand what PII was leaked 
from participant devices but not in our controlled experiments, and to understand users' opinions
about privacy. 

The study includes 92 users in total, with 63 iOS devices and 33 Android devices (some users have more than one device). 
\new{In the initial training phase, we initialized the \recon classifiers with the pre-labeled dataset discussed in  \S\ref{sec:eval}. Then we use the continuous user feedback to retrain the classifiers.} \revise{The anonymized results of PII leaks discovered from our ongoing user study can be found at 
 \begin{small}\url{http://recon.meddle.mobi/app-report.html}\end{small}.}

\noindent\textbf{Runtime.}
While the previous section focused on runtime in terms of training time, an important 
goal for \recon is to predict and extract PII in-band with network flows so that we can 
block/modify the PII as requested by users. As a result, the network delay experienced 
by \recon traffic depends on the efficiency of the classifier.

We evaluated \recon performance in terms of PII prediction and extraction times. The combined 
cost of these steps is less than 0.25\,ms per flow on average (std. dev. 0.88), and never exceeds 6.47\,ms per flow.
We believe this is sufficiently small compared to end-to-end delays \new{of} \revise{tens or hundreds} of milliseconds in mobile \new{networks}.

\noindent\textbf{Accuracy ``in the wild.''}
Participants were asked to view their PII leaks via the \recon UI, and label them as correct or incorrect. 
As of Dec 8, 2015, our study covers 1,120,278 flows, 9,573 of which contained PII leaks that \recon identified. Of those, there are 5,351 TP 
leaks, 39 FP leaks and 4,183 unlabeled leaks. Table~\ref{table-uleaks-os} shows the results across all users. \emph{The users in the study found few cases when \recon 
incorrectly labeled PII leaks.} The vast majority \revise{(85.6\%)} of unlabeled data is device identifiers, \new{likely because it is difficult for users to find such identifiers to compare with our results.}

\noindent\textbf{Impact of user feedback on accuracy.}
\new{To evaluate the impact of retraining classifiers based on user feedback, we compare the results without user feedback (using our 
initial training set only) with those that incorporate user feedback. After retraining the classifier, \emph{the false positive 
rate decreased by 92\% (from 39 to 3), with a minor impact on false negatives (0.5\% increase, or 18/5,351)}. }

\revise{\noindent\textbf{Retraining classifiers.} As discussed in \S\ref{sec:pdao}, we retrain \recon classifiers periodically and after collecting sufficient samples. 
We provide options to set the frequency of retraining and the retraining process is relatively low cost. In our experience, retraining the general classifier once a day or once a week is sufficient to retain high accuracy. This is a process that occurs in the background, takes little time per domain (0.9\,s per domain 
on average), and is easily parallelized to reduce retraining time.}

\noindent\textbf{User survey.}
To qualitatively answer whether \recon is effective, we conducted a survey where we asked participants,
 ``Have you changed your ways of using your smartphone and its applications based on the information provided by our system?''
 Of those who responded to the voluntary survey, a majority (20/26) indicated that they found the system useful and changed their habits related to privacy when using 
 mobile devices. This is in line with results from Balebako et al.~\cite{balebako:soups13}, who found 
 that users ``do care about applications that share privacy-sensitive information with third parties, and would want more information about data sharing.''

\revise{In terms of overhead, we found that a large majority of users (19/26) observed that battery consumption and Internet speed 
were the same better when using \recon. While the remaining users observed increased battery consumption and/or believed their 
Internet connections were slower, we do not have sufficient data to validate whether this was due to \recon or other factors such 
inherent network variations or increased user awareness of these issues due to our question.  }

\noindent\textbf{PII leak characterization.}
We now investigate the PII leaked in the user study. As Table~\ref{table-uleaks-os} shows, the most 
commonly leaked PII is device identifiers, likely used by advertising and analytic services. The 
next most common leak is location, which typically occurs for apps that customize their behavior 
based on user location. We also find user identifiers commonly being leaked (\eg name and gender), 
suggesting a deeper level of tracking than anonymous device identifiers. Depressingly, even in our 
small user study we found 171 cases of credentials being leaked in plaintext (102 verified by users). 
\new{For example, the Epocrates iOS app (used by more than 1 million physicians and health professionals) and 
the popular dating app Match.com (used by millions, both Android and iOS were affected) leaked user credentials in plaintext. Following 
responsible disclosure principles, we notified the app developers. The Epocrates app was fixed as of November, 2015 (and 
the vulnerability was made public~\cite{epocrates-msg} after we gave them time to reach out to users to convince them to upgrade), 
and} \revise{Match.com fixed their password exposure in January, 2016 without notifying us or the public.}
These results highlight 
the negative impact of closed mobile systems---even basic security is often violated by sending passwords 
in plaintext (\revise{21} apps in our study). 

We further investigate the leaks according to OS (Table~\ref{table-uleaks-os}).\footnote{\revise{Note that these results are purely observational and 
we do not claim any representativeness. However, we did normalize our results according to the number of users per OS.}} We find that the average iOS user 
in our study experienced more data leaks than the average Android user, and particularly experienced higher 
relative rates of device identifier, location, and credential leaks. 

\end{sloppypar}

\begin{table}
\centering
\addtolength{\tabcolsep}{-1pt}
\begin{small}
\begin{tabular}{ll|c|c|c|c}
         &                 &                  & \multicolumn{3}{c}{\sl Feedback on leaks}              \tabularnewline 
         & {\bf Leak Type}      & {\bf total}  & {\bf correct}   & {\bf wrong}  & {\bf no label} \tabularnewline
	 &  &   & & & {\bf /unknown} \tabularnewline
\hline
\multirow{5}{*}{\rotatebox{90}{\bf iOS}}
&{ Device ID. } & 3229 & 12 & 35 & 3182 \tabularnewline
&{ User ID. } & 655 & 216 & 2 & 437 \tabularnewline
&{ Contact Info. } & 6 & 3 & 1 & 2 \tabularnewline
&{ Location } & 4836 & 4751 & 0 & 85 \tabularnewline
&{ Credential } & 36 & 30 & 0 & 6 \tabularnewline
\hline
\multirow{5}{*}{\rotatebox{90}{\bf Android}}
 &{ Device ID. } & 399 & 2 & 0 & 397 \tabularnewline
&{ User ID. } & 31 & 30 & 0 & 1 \tabularnewline
&{ Contact Info. } & 8 & 8 & 0 & 0 \tabularnewline
&{ Location } & 238 & 227 & 0 & 11 \tabularnewline
&{ Credential } & 135 & 72 & 1 & 62 \tabularnewline
\end{tabular}
\end{small}
\caption{{\bf Summary of leaks predicted by OS.}
{\sl We observe a higher number of leaks for iOS because the number of iOS devices (63) is more than the number of Android devices (33). } }
\label{table-uleaks-os}
\end{table}

We investigated the above leaks to identify several apps responsible 
for ``suspicious'' leaks. 
For example, the ABC Player app is inferring and transmitting the user's gender. 
Last, All Recipes---a cookbook app---is tracking user locations 
even when there is no obvious reason for it to do so. 

%% file: related.tex
\section{Related Work}
\label{sec:related}
Our work builds upon and complements a series of related work on privacy and tracking. 
Early work focused on tracking via Web browsers~\cite{lightbeam,roesner:webtrackers}.
Mobile devices make significant PII available to apps, and early studies showed PII such as 
location, usernames, passwords and phone numbers were leaked by popular apps~\cite{wtk}. 
Several efforts systematically identify PII leaks from mobile devices, and develop defenses against them.

\noindent\textbf{Dynamic analysis.} One approach, dynamic taint tracking, modifies the device OS to track access to PII at runtime~\cite{enck:taintdroid} using dynamic information flow analysis, which taints PII as it is copied, mutated and exfiltrated by apps. This ensures that all access to PII being tracked by the OS is flagged; however, it can result in large false positive rates (due to coarse-granularity tainting), false negatives (\eg because the OS does not store leaked PII such as a user's password), and incur significant runtime overheads that discourage widespread use. Running taint tracking today requires rooting the device, which is typically conducted only by advanced users, and can void the owner's warranty. Other approaches that instrument apps with taint tracking code still either require modifications to platform libraries~\cite{phosphor}, and thus rooting, or resigning the app under analysis~\cite{uranine}, essentially breaking Android's app update and resource sharing mechanisms. \new{When taint tracking is performed as part of an automated analysis environment, user input generation is crucial to improve coverage of leaks. Tools such as Dynodroid~\cite{Machiry:dynodroid}, PUMA~\cite{Hao:2014:puma}, and A3E~\cite{Azim:2013:a3e} automatically generate UI events to explore UI states, but require manual input for more complex user interactions, \eg logging in to sites~\cite{choudhary2015automated}.} Finally, taint tracking does not address the problem of which PII leaks should be blocked (and how), a problem that is difficult to address in practice~\cite{hornyack:appfence}. \new{Nevertheless, automated dynamic analysis approaches are complementary to \recon: as we demonstrated in \S\ref{subsec:compareTD}, \recon can learn from PII leaks identified through dynamic information flow analysis.}

\noindent\textbf{Static analysis.}  Another approach is to perform static analysis (\eg using data flow analysis or symbolic execution) 
to determine \emph{a priori} whether an app will leak privacy information~\cite{yang:appintent,jeon:symdroid,arzt:flowdroid,egele:pios,appaudit,lu:chex,gibler:androidleaks,kim:scandal,yan:droidscope,zhang:vetting,agarwal:pmp,droidjust}. This approach can avoid run-time overhead by performing analysis before code is executed, but state-of-the-art tools suffer from imprecision~\cite{EdgeMiner} and symbolic execution can be too time-intensive to be practical. Further, deploying this solution generally requires an app store to support the analysis, make decisions about which kinds of leaks are problematic, and work with developers to address them. 
Static analysis is also limited by code obfuscation, and tends not to handle reflection and dynamically loaded code~\cite{stadyna}. A recent study~\cite{andrubis:badgers} finds dynamically loaded code is increasingly common, comprising almost 30\% of goodware app code loaded at runtime. 

\revise{\noindent\textbf{New execution model.} Privacy capsules~\cite{priv-capsules} (PC) are an OS abstraction that prevent privacy 
leaks by ensuring that an app cannot access untrusted devices (\eg a network interface) after it accesses private information, unless 
the user explicitly authorizes it. The authors show the approach is low cost and effective for some apps, but it is currently deployed 
only as a prototype extension to Android and requires app modifications for compliance.  }

\noindent\textbf{Network flow analysis.}
\recon analyzes network flows to identify PII leaks. 
Previous studies using network traces gathered inside a mobile network~\cite{gill2013follow,vallina-rod:ads}, in an ISP~\cite{liu-2015-pii}, and in a lab setting~\cite{bala:diffusion} identified significant tracking, despite not having access to software instrumentation. 
In this work, we build on these observations to both identify how users' privacy is violated and control these privacy leaks \emph{regardless of the device OS or network being used}.

PrivacyGuard~\cite{privacyguard}, AntMonitor~\cite{antmonitor} and \new{HayStack~\cite{razaghpanah:2015:haystack}} use the Android VPNService to intercept traffic and perform traffic analysis. 
A limitation of these approaches is they rely on hard-coded identifiers for PII, or require knowledge of a user's PII to work. 
Further, these approaches \new{currently} work only for the Android OS. 
In contrast, \recon is cross-platform, does not require a priori knowledge of PII, and is adaptive to changes in how PII leaks.

%% file: conclusion.tex
\section{Conclusion}
\label{sec:conc}

In this paper we presented \recon, a system that improves visibility and control over privacy leaks 
in traffic from mobile devices. We argued that since PII leaks occur over the network, 
detecting these leaks at the network layer admits an immediately deployable and cross-platform solution to the problem. Our approach based on machine learning has good 
accuracy and low overhead, and adapts to feedback from users and other sources of ground-truth 
information.  

We believe that this approach opens a new avenue for research on privacy systems, and provides 
 opportunities to improve privacy for average users. We are investigating 
how to use \recon to build a system to provide properties such as k-anonymity, or allow users to 
explicitly control how much of their PII is shared with third parties---potentially doing so in 
exchange for micropayments or access to app features.

\section{Acknowledgements}
\revise{We thank the anonymous reviewers and our shepherd, Ben Greenstein, for their feedback. We also thank our study participants, and for contributions toward early work in this area from Justine Sherry, Amy Tang, and Shen Wang. 

The \recon project is supported by the Data Transparency Lab. 
The research leading to these results has 
also received funding from the FFG~-- Austrian Research Promotion under grant COMET K1 
and from u'smile, the Josef Ressel Center 
for User-Friendly Secure Mobile Environments. 
Ashwin Rao was partially supported by a research grant from Nokia.}